\documentclass[pre,aps,superscriptaddress,showpacs,preprintnumbers]{revtex4}
\usepackage{epsfig} 
\usepackage[latin1]{inputenc}
\usepackage{amsfonts}
\usepackage{amsmath}

\newcommand{\imag}{\hat{\imath}}
\newcommand{\dd}{\mathrm{d}}
\newcommand{\id}{1\hspace{-2.6pt}\mathrm{I}}
\newcommand{\phib}{\overline{\phi}}
\newcommand{\psib}{\overline{\psi}}
\newcommand{\chib}{\overline{\chi}}
\newcommand{\xib}{\overline{\xi}}
\newcommand{\phit}{\tilde{\phi}}
\newcommand{\chit}{\tilde{\chi}}
\newcommand{\rhoi}{\rho_0}
\newcommand{\rhof}{\rho_T}
\newcommand{\nui}{\nu_0}
\newcommand{\nuf}{\nu_T}
\newcommand{\lambdac}{\lambda_\mathrm{C}}

\newcommand{\picm}{\pi_\mathrm{MF}}
\newcommand{\eqn}{Eq.~}
\newcommand{\ddt}[1]{\frac{\dd #1}{\dd t}}

\begin{document}

\title{Field-theoretic approach to metastability in the contact process}

\author{Christophe Deroulers}
\email{Christophe.Deroulers@ens.fr}
\affiliation{Laboratoire de Physique Th{\'e}orique de l'ENS, 24 rue 
Lhomond, 75231 Paris CEDEX 05, France}
\author{R\'emi Monasson}
\email{monasson@lpt.ens.fr}
\affiliation{CNRS-Laboratoire de Physique Th{\'e}orique de l'ENS, 24 rue
Lhomond, 75231 Paris CEDEX 05, France}
\affiliation{CNRS-Laboratoire de Physique Th{\'e}orique, 
3 rue de l'Universit\'e, 67000 Strasbourg, France}

\date{\today} 

\begin{abstract}
A `quantum' field-theoretic formulation of the dynamics of the Contact
Process on a regular graph of degree $z$ is introduced. A perturbative
calculation in powers of $1/z$ of the effective potential for the density
of particles $\phi(t)$ and an instantonic field $\psi(t)$ emerging from
the formalism is performed.  Corrections to the mean-field
distribution of densities of particles in the out-of-equilibrium
stationary state are derived in powers of $1/z$. Results for typical (e.g.
average density) and rare fluctuation (e.g. lifetime of the metastable
state) properties are in very good agreement with numerical simulations
carried out on $D$-dimensional hypercubic ($z=2\,D$) and Cayley lattices.
\end{abstract}

\preprint{LPTENS 03/30}

\pacs{05.70.Ln, 02.50.-r, 03.50.Kk, 05.10.-a, 05.40.-a, 05.50.+q, 
64.60.Cn, 64.60.My, 75.10.Hk, 82.20.-w}

\maketitle

\section{Introduction}

\subsection{Motivations}

Recent years have seen an upsurge of interest for the dynamical
properties of out-of-equilibrium systems in statistical physics
\cite{marro-dickmann}. Systems 
of interacting elements are ubiquitous in physics and other fields,
e.g. biology, computer science, economy... Most of the time the 
dynamical rules do not obey detailed balance or similar criteria
which would ensure the existence of a well defined stationary distribution 
at large times. In other cases, a Gibbs measure does exist but is 
out-of-reach on experimental time scales, and all phenomena of interest 
e.g. the occurrence of dynamical phase transitions take place when 
the system is truly out-of-equilibrium. An example of such
out-of-equilibrium phenomena, frequently encountered in condensed
matter, in cellular automata or even in computationally-motivated 
problems is the occurrence of metastable states, or regions in phase 
space in which trapping may take place for a very long time before 
further evolution becomes possible.

The calculation of the temporal properties of these systems often turns
out to be very hard, even when dynamical rules look like innocuously
simple. Over the past decade, however, various models and problems have
been successfully investigated e.g.
\cite{henkel-equivalences-reaction-diffusion-integrables-1994,
henkel-equivalences-reaction-diffusion-integrables-1997,
derrida-lebowitz-asep-fonction-grande-deviation-exacte,
derrida-revue-asep, persistence-majumdar-1999}. Among the analytical
methods used to tackle these systems, some rely on the observation that
the master equation for a system of classical degrees of freedom may be
seen as a Schr\"odinger equation (in imaginary time) where the quantum
Hamiltonian encodes the evolution operator. Exact e.g. Bethe Ansatz
\cite{henkel-equivalences-reaction-diffusion-integrables-1997} or
approximate e.g. variational or semi-classical techniques developed in the
context of quantum field theory may be used to understand the dynamical
properties of the original system \cite{hinrichsen-revue-2000}. One 
important achievement made possible by this approach once combined with
renormalization group techniques has been the calculation of decay
exponents and the identification of universality classes in
reaction-diffusion models \cite{cardy-livre-1996, cardy-notes}.

The range of this ``quantum'' procedure is however not limited 
to the calculation of universal quantities. In this work, we show
how it can be combined to diagrammatic techniques developed in the 
contexts of field theory and the statistical physics of
disordered systems to quantitatively characterize 
the metastable properties of a well-known example of out-of-equilibrium 
system, the so-called contact process (CP)\cite{definition-proccont-harris}. 
In spite of its technicalities, this approach allows us to make predictions
that can be successfully compared to numerical simulations. It is
expected that it will permit to investigate metastability 
\cite{metastabilite-proccont-schonman, metastabilite-bootstrap-lebowitz}, 
or other properties of various dynamical models.

\subsection{The Contact Process: Definition and Phenomenology}
\label{secpheno}

\begin{figure}
\begin{center}
A \epsfig{file=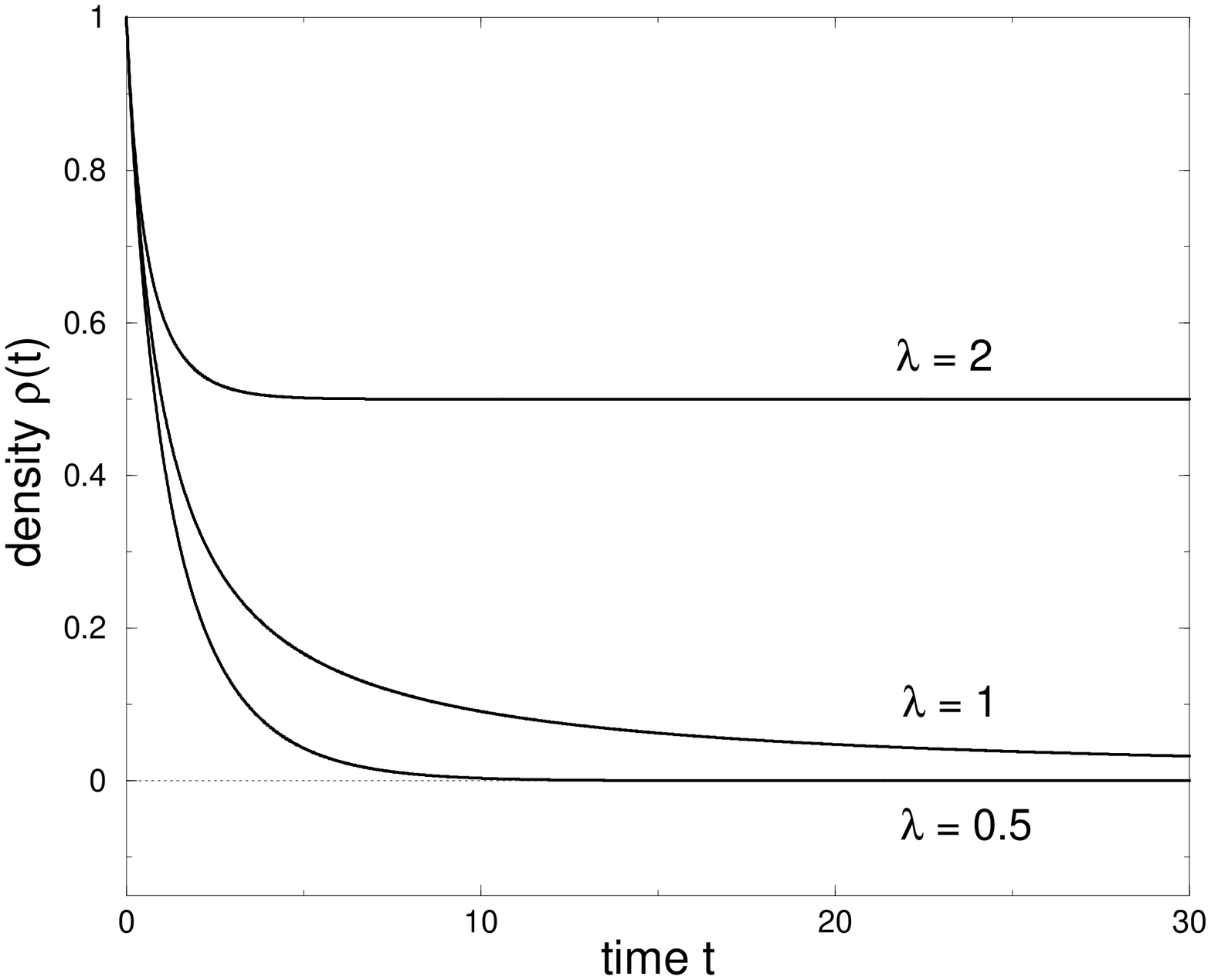,width=6cm,angle=-90}
\hskip 1cm
B \epsfig{file=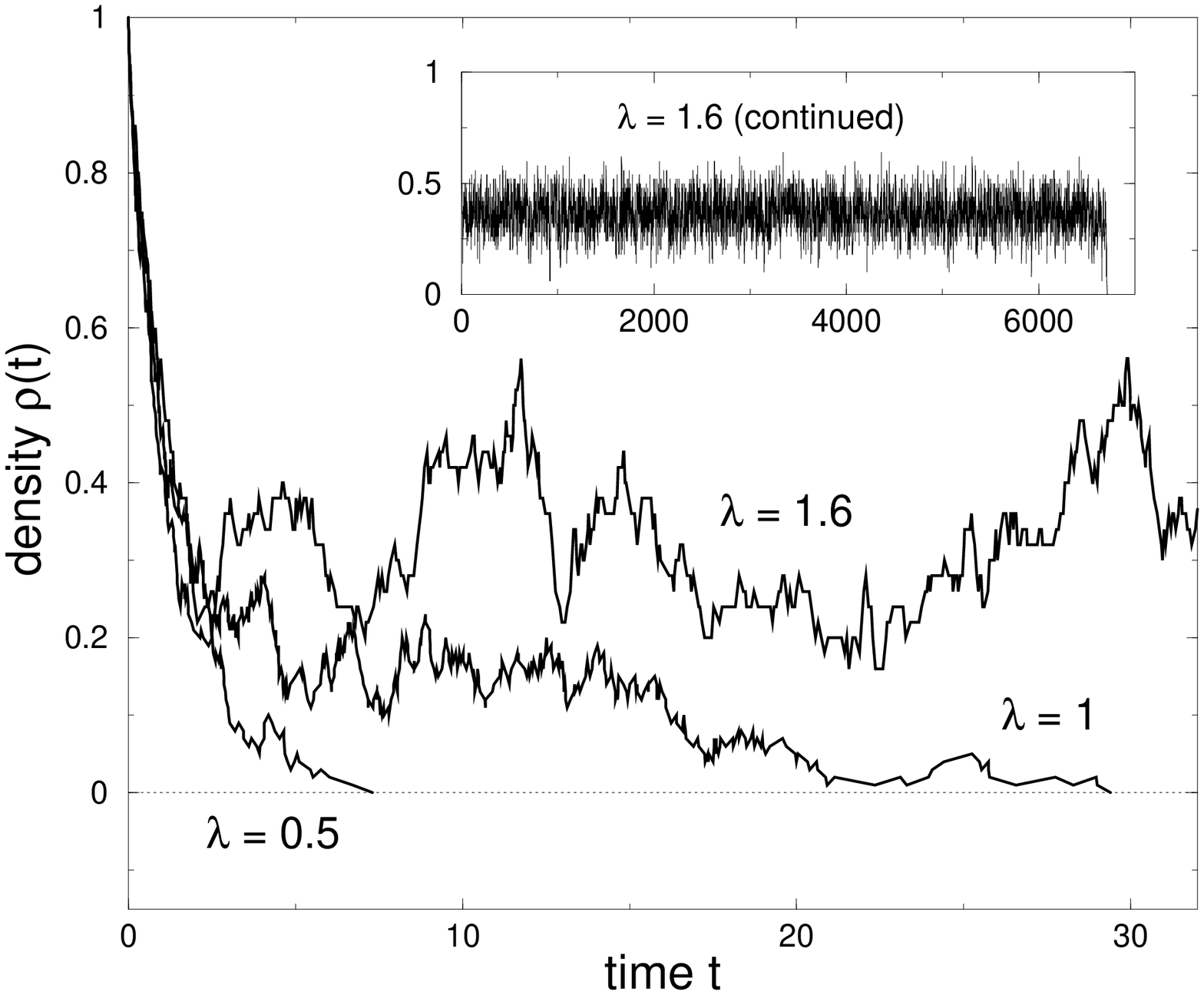,width=6cm,angle=-90}
\end{center}
\caption{Profiles of the density $\rho$ of particles versus time $t$ 
for the Contact Process over a complete graph of $N$ vertices, 
initially filled with particles ($\rho (0)=1$). 
{\bf A.} Thermodynamic limit, $N\to \infty$. 
From bottom to top: subcritical ($\lambda <\lambdac = 1$, the
density exponentially relaxes toward zero), critical ($\lambda =\lambdac$,
the density algebraically decays to zero as $\rho (t) \sim t^{-1}$), and 
supercritical ($\lambda > \lambdac$, the density exponentially relaxes 
to a finite value, $\rho ^* = 1-1/\lambda$) cases. 
The density obeys a deterministic evolution equation
(\ref{dyn56}), and no fluctuation is present. {\bf B.} Finite size lattice,
with $N=100$ sites. CP is a stochastic process, and the density profiles 
vary from run to run (we show here one run for each value of
$\lambda$). The density quickly relaxes to zero (subcritical regime)
or a finite value (supercritical regime). In the latter case, 
the system is trapped in a metastable regime where the density
fluctuates for a very long time around its plateau value (Inset, notice 
the difference of time scale), till a large fluctuation 
drives the density to the zero value.}
\label{figures-densite}
\end{figure}

We consider a regular graph $G$ with vertex degree $z$ and size $N$
(number of vertices). Each vertex (or node, or site) may be empty or
occupied by one particle.  Hereafter, we focus on the continuous time
version of CP where a particle is spontaneously
annihilated with rate 1 independently from other sites, and an
empty site becomes occupied with rate $\lambda \, n_{\mathrm{occ}} /
z$ where $n_{\mathrm{occ}}$ is the number of its occupied nearest
neighbours \cite{definition-proccont-harris}.

The value of the parameter $\lambda$ strongly affects the behaviour of
CP.  For infinite size graphs e.g. infinite hypercubic lattices, there 
exists a critical value $\lambdac$ of $\lambda$ such
that \cite{liggett-livre-1985, liggett-livre-1999},
\begin{itemize}
\item if $\lambda < \lambdac$, the number of particles (occupied
sites) quickly decreases towards zero. Later, the system remains
trapped in this empty configuration.
\item if $\lambda > \lambdac$, the density $\rho$ of particles
reaches a plateau value $\rho ^*(\lambda)$ independently on the
initial density \footnote{More precisely, the probability that the
plateau is reached tends to one as $N\to\infty$ for any initial
configuration with density $\epsilon >0$.}.
\item at criticality, that is, when $\lambda = \lambdac$, the density
eventually relaxes to zero with a slow algebraic decay $\rho (t) \sim
t^{-a}$. This critical behaviour falls into the directed percolation
universality class~\cite{grassberger-reactiondiffusion-reggeon-1978,
cardy-reactiondiffusion-reggeon-1980,
janssen-reactiondiffusion-reggeon-1981}. Exponent $a$ is equal to $1$ in
dimensions larger than $D_c=4$, and approximate expressions in powers of
$D_c-D$ in lower dimensions have been obtained through the use of
renormalization group
techniques~\cite{grassberger-reactiondiffusion-reggeon-1978,
cardy-reactiondiffusion-reggeon-1980,
janssen-reactiondiffusion-reggeon-1981,
grassberger-reactiondiffusion-reggeon-1982, hinrichsen-revue-2000,
cardy-notes, cardy-livre-1996}.
\end{itemize}
These behaviors are displayed in Fig.~\ref{figures-densite}A.
The exact value of the critical parameter $\lambdac$ is unknown in any
dimension $D$, but rigorous bounds and estimates have been derived 
\cite{liggett-livre-1999}.

\begin{center}
\begin{figure}
\epsfig{file=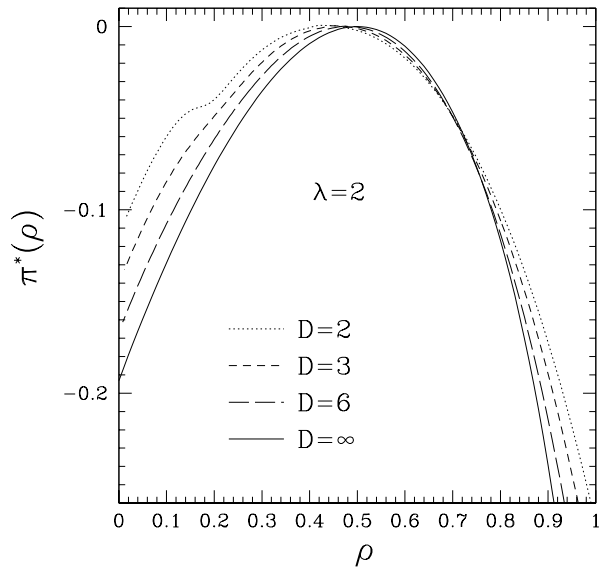,width=8cm}
\caption{The large deviation function $\pi^*$ for the density of particles
in CP with parameter $\lambda=2$ for the $D$-dimensional hypercubic 
lattice. The $D\to\infty$ curve corresponds to the mean-field limit, and 
coincides with the case of the complete graph over $N\to\infty$ sites.
Plot of the predicted value of $\pi^*(\rho)$ at
the order $1/D$ for $D= 6, 3$ and $2$ are obtained with the expansion
of Section IV. The nonconvexity of the curve for $D=2$ shows the 
inaccuracy of the truncation to first order of our $1/D$ expansion in this 
range of densities.}
\label{courbes_pi_de_rho_predites}
\end{figure}
\end{center}

For finite size graphs, the empty configuration, referred to as vacuum
in the following, is an absorbing state for the dynamics. Starting
from any initial configuration e.g. fully occupied state, CP will
eventually end up in the vacuum configuration after a finite time
$t_{vac}$. This forces the above infinite size picture to be smeared
out by fluctuations in the case of large but finite lattices
\cite{metastabilite-proccont-schonman, 
grandesdeviations-proccont-durett-schonman, liggett-livre-1999}. 
$\lambdac$ locates a cross-over between fast ($t_{vac} (N, \lambda < 
\lambdac) = O(\log N)$) and very slow ($t_{vac} (N, \lambda > \lambdac) 
\sim \exp O(N)$) relaxations towards the vacuum configuration. In the 
latter regime, the plateau height, $\rho ^*$, merely defines an average 
value around which the density exhibits fluctuations until the system is
driven to the vacuum through a very large fluctuation
(Fig. \ref{figures-densite}B).  On time scales
$1 \ll t \ll t_{vac} (N, \lambda > \lambdac)$, the system is trapped
into a metastable state~\cite{simonis-proccont-metastable}. A
(pseudo-)equilibrium probability measure for the density can be defined,
\begin{equation} \label{defpi}
P(\rho , N) = \exp \big(  N \, \pi^* (\rho ) + o(N) \big) \quad .
\end{equation}
Function $\pi^*$, which depends on $\lambda$ and other parameters e.g. the 
dimension $D$ for hypercubic lattices, describes rare fluctuations
from the average density $\rho ^*$. Its maximal value is zero for
$\rho = \rho ^*$. Densities $\rho$ distinct from the average one are
exponentially (in $N$) unlikely to be reached, and $\pi^* (\rho ) <
0$, see Fig. \ref{courbes_pi_de_rho_predites}. 
In particular, the probability of a very large fluctuation
annihilating all particles scales as $\exp (  N \pi^* (0))$, and thus,
\begin{equation} \label{ret}
t_{vac} (N) = \exp \big( - N \pi ^*(0) + o(N) \big) \quad .
\end{equation}
The calculation of the large deviation function $\pi^*(\rho)$ is the main scope
of this paper. For this purpose we use a path integral representation of 
$\pi^*$ where particles are encoded into quantum hard core bosons, or 
$1/2$-spins, and develop a diagrammatic self-consistent
evaluation of the path integral which allows us to write a systematic 
expansion of $\pi^*$ in powers of $1/z$ (Section II). In the infinite
connectivity limit ($z\to\infty$), this formalism reduces to the 
mean-field theory of CP, analyzed in Section III. Finite connectivity
corrections to $\pi ^*$ are  calculated in Section IV. 
As a by-product, we obtain an expansion for the critical parameter 
$\lambdac(z)$ in powers of $1/z$. The validity of
our calculation, effectively carried out up to order $1/z^2$ (and $1/z$ 
for $\pi^*$), is confirmed by numerical simulations performed on
$D$-dimensional hypercubic ($z=2D$) and Cayley lattices.

\section{Field-theoretic framework}
\label{secfieldtheory}

\subsection{Path integral formulation of the evolution operator}

We start by writing the master equation of CP using a quantum formalism,
according to the familiar procedure of Felderhof, Doi and
successors~\cite{felderhof, doi}. For each site $i$ of the graph we define
a hard-core boson with associated state vectors $|0\rangle_i$ (empty)  
and $|1\rangle_i$ (occupied), and creation and annihilation operators
$a^+_i$ and $a_i$ that anticommute on a single site but commute on
different sites: $[a_i,a^+_i]_{+} = \id$, $[a^+_i,a^+_j] = [a_i,a^+_j]=
[a_i,a_j]=0$ (alternately, we could use spins $1/2$ with
a mere rewriting of the equations). To each occupation number $s_i=0,1$ of
site $i$ is associated a vector $|s_i\rangle$. Then, to each state $\vec
s$ of the graph (set $(s_1,s_2,\ldots ,s_N)$ of occupation numbers of all
the sites) corresponds \cite{kadanoff-equation-maitresse-quantique} a 
basis vector of a $2^N$-dimensional vector space,
$|\vec s\rangle = |s_1\rangle \otimes |s_2\rangle \otimes \ldots \otimes
|s_N\rangle$, and, to the time-dependent probability
distribution $P(\vec s,t)$, the state vector $|P(t)\rangle = \sum_{\vec 
s} P(\vec s,t) |\vec s\rangle$.
The master equation for $P(\vec s,t)$ is now equivalent to the
evolution equation of the state vector 
\begin{equation}
\ddt{} |P(t)\rangle = \hat W \;|P(t)\rangle
\end{equation}
where the evolution operator $\hat{W}$ is the infinitesimal
generator of the transitions. For CP, $\hat{W}=
\hat{W}_{\mathrm{ann}} + \lambda \hat{W}_{\mathrm{cre}}$ with
\begin{eqnarray}
\hat{W}_{\mathrm{ann}} &=& \sum_i (\id-a^+_i)\; a_i \nonumber \\
\hat{W}_{\mathrm{cre}} &=& \frac 1z \sum_i \sum_{j \in i} 
\big( a^+_j \, (\id + a_j ) - \id \big)\;  a^+_i a_i 
\end{eqnarray}
where $j \in i$ means that site $j$ is one of the $z$ nearest neighbours 
of site $i$. 

To map the stochastic process onto a path integral or field-theoretic
formulation \cite{peliti-integrale-de-chemins, cardy-livre-1996,
cardy-notes}, we
introduce~\cite{klauder-etats-coherents-pour-integrale-de-chemins-1960,
klauder-etats-coherents-pour-integrale-de-chemins-1979, klauder-livre}
continuously parametrized states suitable for hard-core 
bosons~\footnote{Another possibility would be to follow the approach of 
Ref. \cite{vanwijland-integrale-de-chemins}}. On each site of the graph, 
the state bra and ket are, respectively,
\begin{eqnarray} \label{etat-coherent-phitheta}
\langle \phi, \theta | &=& (1-\phi)^\frac{1}{2} \langle 0 | + 
 \phi^\frac{1}{2} \; \exp(-\imag \theta) \; \langle 1 | \quad, \nonumber \\
| \phi, \theta \rangle &=& (1-\phi)^\frac{1}{2} \; |0\rangle +
 \phi^\frac{1}{2} \; \exp(\imag \theta) \; |1\rangle \quad
\end{eqnarray}
where $\phi\in [0,1]$, $\theta \in [0,2\pi]$ and $\imag ^2=-1$. These
states satisfy the closure relation:
\begin{equation}
 \label{relation-fermeture-etat-coherent}
\frac{1}{\pi} \int_0^1 \dd \phi \int_0^{2\pi} \!\! \dd \theta \;
 |\phi, \theta \rangle \langle \phi, \theta | = \id \quad .
\end{equation}
To allow a simplification of the expressions in the translation table 
given below, we make use of
\mbox{$\psi := -\frac{1}{2} \ln[\phi /(1-\phi)] +
\imag \;\theta$} instead of $\theta$, and introduce the following 
notations,
\begin{eqnarray} \label{coh}
\langle \phi, \psi | &=& (1-\phi)^{\frac 12} \; 
\bigg( \langle 0 | + \exp(-\psi) \;\langle 1 | \bigg) \quad, \nonumber \\
| \phi, \psi \rangle &=& (1-\phi)^{-\frac 12} \; \bigg(
(1-\phi) \; |0\rangle + \phi\; \exp(\psi) \;  |1\rangle \bigg)\quad .
\end{eqnarray}
For the whole graph, a state is the tensor product of the states over all
sites: $|\vec{\phi}, \vec{\psi}\rangle = \otimes_{i=1}^N |\phi_i,
\psi_i\rangle$. Making use of Trotter formula \cite{schulman-livre} and of
the closure identity~(\ref{relation-fermeture-etat-coherent}), we obtain a
path-integral expression for the matrix elements of the evolution operator
$\exp (T \, \hat{W})$ between times 0 and $T$
\cite{peliti-integrale-de-chemins, cardy-livre-1996},
\begin{equation}
\label{intchem_phipsi}
\langle \vec{\phi}_T, \vec{\psi}_T | \exp(T\;\hat{W} ) |
  \vec{\phi}_0, \vec{\psi}_0 \rangle =  \int 
_{\vec \phi(0)=\vec \phi_0, \vec \psi(0)=\vec \psi_0}
^{\vec \phi(T)=\vec \phi_T, \vec \psi(T)=\vec \psi_T}
{\cal D}  \vec{\phi}(t)\, {\cal D}\vec{\psi}(t) \; 
\exp \left( - {\cal S} [ \{ \vec{\phi} , \vec{\psi} \} ] \right) \ ,
\end{equation}
where the action reads
\begin{equation} \label{action_def}
-{\cal S} [ \{ \vec{\phi} , \vec{\psi} \} ] 
= - \int_0^T \dd t \; \left\{ \sum_{i=1}^N \phi_i(t)
\ddt{\psi_i(t)} - \tilde{W}(\vec{\phi}(t),\vec{\psi}(t)) 
\right\} \quad ,
\end{equation}
and the integral runs over all field configurations $\vec \phi(t),
\vec \psi(t)$ over the time interval $t\in [0,T]$ matching the required 
boundary conditions at initial and final times.

Function $\tilde W$ encodes the action of the evolution operator $\hat W$
on the states. Its expression is obtained by first writing $\hat{W}$ in
normal order form thanks to the (anti)commutation relations, then using
the translation Table \ref{table-transformation-operateurs-elementaires},
see \cite{peliti-integrale-de-chemins,cardy-livre-1996, cardy-notes}.  
For CP, we obtain $\tilde{W} = \tilde{W}_{\mathrm{ann}} + \lambda
\tilde{W}_{\mathrm{cre}}$ with
\begin{eqnarray} \label{entriescreaanni}
\tilde{W}_{\mathrm{ann}}(\vec{\phi},\vec{\psi}) &=& \sum_{i=1}^N 
\phi_i \big (\exp (\psi_i)-1\big)  \quad , \nonumber \\
\tilde{W}_{\mathrm{cre}}(\vec{\phi},\vec{\psi}) &=& \frac 1z\sum_{i=1}^N 
\phi_i \sum_{j \in i} (1-\phi_j)  \big(\exp(-\psi_j)-1\big) \quad .
\end{eqnarray}

\begin{table}
\begin{center}
  \begin{tabular}{c|c}
  \hline
  operator in $\hat W$ & expression in $\tilde W$   \\
  \hline
   $\id$ & $1$ \\
$a$ & $\phi \; \exp(\psi)$ \\
$a^+$ & $(1-\phi) \; \exp(-\psi)$ \\
$a^+a$ & $\phi$ \\
$a a^+$ & $1-\phi$ \\
   \hline
  \end{tabular}
\caption{Translation table from operators in $\hat W$ into Lagrangian
contribution to $\tilde W$. We have added $a \; a^+$ in the left column 
though this operator is not in normal order to show the consistency of
the translation rules with the anticommutation relation.}
\label{table-transformation-operateurs-elementaires}
\end{center}
\end{table}

The previous quantum formalism allows us to express the expectation value 
of any observable of interest. 
For instance, we may start at time $t=0$ from a random
state $|\rhoi \rangle$ with exactly $N_0 = \rhoi\, N$ occupied sites 
and project at the end on a state $\langle \rhof|$ with exactly $N_T =
\rhof\,N$ occupied sites:
\begin{equation}
|\rhoi \rangle :=  \frac 1{\binom{N}{N_0}} \sum_{\vec s} 
  \otimes_{i=1} ^N \bigg( (1-s_i) \;|0\rangle + s_i\; |1\rangle \bigg)
\quad,\quad
\langle \rhof | :=  \sum_{\vec s} \otimes_{i=1}^N
  \bigg( (1-s_i)\;\langle0| + s_i\;\langle1| \bigg)
\end{equation} 
where the sums run over all states $\vec s$ with $N_0$ (resp. $N_T$)
occupation numbers $s_i$ equal to 1, and the remaining $N-N_0$ (resp. 
$N-N_T$) ones equal to 0. The probability that the density of particles 
equals $\rhof$ at time $t=T$ given that is was equal to $\rhoi$ at time 
$t=0$ is thus $P(\rhof,T |\rhoi,0) =  \langle \rhof| \exp(T \, \hat{W} 
) | \rhoi \rangle$. Using the path integral formalism developed in this 
section, the logarithm $\Pi$ of this probability reads
\begin{equation}
\label{pi_de_rho_integrale_de_chemins}
\Pi(\rhof ,T | \rhoi ,0) = \ln \bigg[
\int {\cal D}  \vec{\phi}(t)\, {\cal D}\vec{\psi}(t) \; 
\exp \bigg( - {\cal S} [ \{ \vec{\phi} , \vec{\psi} \} ] \bigg)
\;\langle \rhof | \vec \phi(T) , \vec \psi (T) \rangle
\; \langle \vec \phi(0) , \vec \psi (0) | \rhoi \rangle \bigg]
\end{equation}
where the boundary conditions for the fields at initial and final times
are now free. Knowledge of the above function gives access to the
large deviations function defined in \eqn (\ref{defpi}) through
\begin{equation}
\pi ^* (\rho ) = \lim_{T\to\infty} \lim_{N\to\infty} \frac 1N  \;
\Pi(\rho  ,T | \rho ^* ,0) 
\quad .
\end{equation}
Notice that, although there is no notion of energy nor Hamiltonian in CP, 
the form of its path-integral formulation closely looks like a classical 
mechanics Lagrangian, with a kinetic energy term and an effective 
potential energy term.
 
Calculation of the path integral on the r.h.s. of \eqn 
(\ref{pi_de_rho_integrale_de_chemins}) will be done through
a perturbative expansion (in $\lambda$) of the effective potential
for the average values of the fields $\vec \phi(t), \vec\psi (t)$, 
following an approach 
used in the context of classical statistical mechanics 
\cite{series-gaunt-baker,agjy-ising}. This expansion
allows us to calculate quantities of interest e.g. $\phi ^* (\rho)$, 
$\lambdac$, $\rho^*$,~... in powers of $1/D$. In the
following, we will closely follow the technique and notations
of Ref. \cite{agjy-ising} which makes use of this perturbation expansion
scheme to calculate equilibrium properties of the Ising model in large
dimensions. The main difference (and complication) is that, here, 
fields depend on time. An application of this approach to the study
of the dynamics of continuous spins models can be found in \cite{giulio}.

\subsection{Diagrammatic expansion of the effective potential}

\subsubsection{Constrained fields and conjugated sources.}

Let $\phib_i(t)$ and $\psib_i(t)$ be two arbitrary 
functions depending on time $t$ and site $i$, with $\phib_i \in [0,1]$,
from which we define $\chib_i(t) := (1-\phib_i(t)) (\exp(-\psib_i(t)) -
1)$. We choose as elementary (site-attached) operators 
$\id -\hat{\phi_i} := \id - a^+_i a_i$ and $\hat{\chi_i} := a^+_i
(\id + a_i) - \id$~\footnote{A natural choice would
have been $\langle a^+_ia_i(t) \rangle = \phib_i(t)$ and $\langle a_i(t)
\rangle = \phib_i(t) \exp(\psib_i(t))$. However this choice gives rise
to a more complicated combinatorics in the $1/D$ expansion. $\hat{\chi}$,
which appears aside $a^+ a$ on a different graph site in the expression 
of $\hat{W}$ is, in this respect, a better `elementary' operator.
Furthermore, we will see that the expectation values of the other 
elementary operators evaluated with this choice will be consistent.
}, and impose the constraints
\begin{equation} \label{constraint_field}
\langle \id - \hat \phi_i (t) \rangle = 1 - \phib_i(t) \quad , \qquad
\langle \hat{\chi_i}(t) \rangle = \chib_i(t) \quad ,
\end{equation}
where $\langle \hat A \rangle=\langle \rhof | \exp(\int_t^T \dd t'
\hat W) \, \hat A\, \exp(\int_0^t \dd t' \, \hat W) |\rhoi\rangle /
\langle \rhof | \exp(\int_0^T \dd t' \hat W) |\rhoi\rangle$
is the average value of operator $A$ at time~$t$. This can be done 
through the introduction of Lagrange multipliers (sources)
in the evolution operator: $\hat W$ is changed to $\hat{W'}(t)
+ W''(t) \; \id$ in the definition of $\langle \hat{A} \rangle$ where
\begin{eqnarray}
\label{definition_W_biaise}
\hat{W'}(t) &:=& \mu\; \hat{W}_{\mathrm{ann}} + \lambda 
\;\hat{W}_{\mathrm{cre}}
 - \sum_i \bigg[ h_i(t)  \; (\id - \hat \phi_i ) 
 - g_i(t) \; \hat{\chi_i}  \bigg] \quad , \nonumber \\ 
W''(t) &:=& \sum_i \bigg[  h_i(t)  \; (1 - \phib_i(t))
 - g_i(t) \;\chib_i(t) \bigg] \quad .
\end{eqnarray}
Fields $h_i(t)$ and $g_i(t)$ are expected to be as regular as the 
imposed order parameters $\phi_i(t)$ and $\chi_i(t)$, and are
assumed to be (at least) once differentiable with continuous 
derivatives over the time interval $t\in ]0;T[$. However, to match with 
the components of the final bra and initial ket, Dirac's 
$\delta$-singularities may be present 
at $t=0$ and $t=T$. Note the introduction of a new 
parameter, $\mu$, in front of the annihilation operator in the expression 
of $\hat W'$. This parameter will result convenient for technical reasons 
only, and we will ultimately be interested in calculating quantities for 
$\mu=1$. This biased evolution operator allows us to express the logarithm 
of the probability that the final density equals $\rhof$ for a fixed set 
of order parameters,
\begin{equation}
\label{definition_pi_de_rho_biaise_canonique}
\Pi \big[\rhof,T ; \{\phib,\chib\} | \rhoi,0\big] = 
 \ln\ \langle \rhof| \hat{\mathcal{W}}(T,0) | \rhoi \rangle +
 \int_0^T \dd t \;W''(t) \quad ,
\end{equation}
and our task will be to compute $\hat{\mathcal{W}}(T,0) := \exp( \int_0^T
\dd t \hat{W'}(t) )$. Requiring that
(\ref{definition_pi_de_rho_biaise_canonique}) be extremal with respect to
$\phib_i(t)$ and $\chib_i(t)$ in addition to the constraints above ensures
that $h_i(t)=g_i(t)=0$ at the extremum of $\Pi$. 
Therefore, at the saddle point, $\Pi$ in \eqn 
(\ref{definition_pi_de_rho_biaise_canonique}) will coincide with $\Pi$ 
defined in \eqn (\ref{pi_de_rho_integrale_de_chemins}). 

The effective potential $\Pi$ can be expanded in a double power series
in $\lambda$, $\mu$, 
\begin{equation}
 \label{double_developpement_Pi}
\Pi = \sum_{a,b\ge 0} \lambda ^a \; \mu ^b \; \Pi_{a,b} \qquad 
\hbox{\rm with} \qquad  \Pi_{a,b}  := \frac 1{a!}\; \frac 1{b!}
(\partial_\lambda) ^a\; (\partial_\mu) ^b\; \Pi|_{\lambda=\mu=0} \quad . 
\end{equation}
We calculate below $\Pi_{0,0}$, that is, the effective potential
in the absence of any evolution process albeit the one resulting from
the kinetic constraint over the order parameters, and then expose how 
to obtain higher orders in $a,b$ through a systematic diagrammatic 
expansion. A nice feature of this expansion scheme is that, at any
given order $a$ in $\lambda$, we are able to resum the whole series 
in powers of $\mu$ and, thus, to express our result as a unique 
power series,
\begin{equation}
\Pi = \sum_{a \ge 0} \lambda ^a  \; \Pi_{a} (\mu) \qquad 
\hbox{\rm with} \qquad  \Pi_{a}  := \sum_{b\ge 0}\; \mu ^b 
\; \Pi_{a,b} \quad , 
\end{equation}
and set $\mu=1$ in the above expression.

\subsubsection{Calculation of $\Pi_{0,0}.$}

We set $\lambda=0$. $\hat{W'}$ decouples into a tensor product over the 
sites. The latter remain however coupled by the constraints that the bra
$\langle \rhof|$ and the ket $|\rhoi \rangle$ correspond to configurations
including exactly  $N_T=\rhof N$ and $N_0=\rhoi N$ particles 
respectively. We thus introduce two further Lagrange multipliers, $\nuf$ 
and $\nui$, to select the initial and final densities of particles. We 
replace $\langle \rhof |$ and $| \rhoi \rangle$ in \eqn
(\ref{definition_pi_de_rho_biaise_canonique}) with, respectively, $\langle
\nuf, \rhof | := \langle O | \exp \big[ \nuf \sum_i (\rhof - 
a^+_i a_i) \big]$ and $| \nui, \rhoi \rangle := 
\binom{N}{N_0}^{-1} \exp \big[
\nui \sum_i (a^+_i a_i - \rhoi) \big] | O \rangle$ 
where $| O\rangle = \left( |0\rangle  + |1 \rangle 
\right) ^{\otimes N} $ is the sum of all possible configurations.
Note that $| \nui, \rhoi \rangle$ is
normalized so as to represent a probability distribution.
Once sites are decoupled, $\Pi$ may be expressed as a sum of site-dependent
effective potentials, each depending upon $\phib_i (t)$, 
$\chib_i (t)$, $h_i(t)$, $g_i (t)$, $\nuf$ and $\nui$. 
We will eventually optimize the resulting $\Pi$ over $\nuf$ and 
$\nui$ to ensure that the final and initial densities are the 
requested ones. 

We send $\mu$ to zero to make $\hat{W'}$ diagonal in the basis
$(|0\rangle,|1\rangle)$. This allows us to compute exactly the evolution
operator $\hat{\mathcal{W}}(t_2,t_1) := \exp( \int_{t_1}^{t_2} \hat{W'}(t)
\dd t)$,  
and then any average of operators or correlation function (CF)  
e.g. $\langle a_i(t_2) a^+_i(t_1)  \rangle := \langle \nuf, \rhof |
\hat{\mathcal{W}}(T,t_2)\; a_i\; \hat{\mathcal{W}}(t_2,t_1)\;  a^+_i
\;\hat{\mathcal{W}}(t_1,0) | \nui, \rhoi \rangle$. 
Evaluating $\langle
(\id - \hat \phi_i)(t)  \rangle$ and $\langle \hat{\chi_i} \rangle$, and
imposing constraints (\ref{constraint_field}), we find back the rules
listed in Table \ref{table-transformation-operateurs-elementaires} with 
over-barred fields \footnote{This consistency statement holds also for 
other choices of Lagrange multipliers in $\hat{W'}$.}. 
The expressions for the sources $h_i(t)$ and $g_i(t)$ are 
then~\footnote{The same expressions are also recovered if one optimizes 
directly $\Pi_{0,0}$ with respect to $h_i(t)$ and $g_i(t)$.}, for times 
$0<t<T$,
\begin{eqnarray}
h_i(t) &=& \ddt{} \psib_i(t) + \big( e^{\psib_i (t)} -1\big)\;
\ddt{} \ln \big( 1- \phib_i (t) \big) \quad , \nonumber \\
g_i (t) &=& - e^{\psib_i (t)} \;
\ddt{} \ln \big( 1- \phib_i (t) \big) \quad ,
\end{eqnarray}
from which we obtain $h_i(t) (1-\phib_i(t)) - g_i(t) \chib_i(t) =
(1 - \phib_i(t)) \ddt{} \psib_i(t)$. 
In other words, the term in $W''$ of \eqn
(\ref{definition_pi_de_rho_biaise_canonique}) gives back, aside boundary
terms involving initial and final values, the `kinetic' term of the
action ${\cal S}$ (\ref{action_def}).

It appears that the constraints (\ref{constraint_field}) at times $t=0$ 
and $t=T$ can be imposed by nonsingular sources $h_i(t)$ and $g_i(t)$ 
only if the required state vectors $\langle \phib_i(T) , \psib_i(T)|$ 
and $|\phib_i(0) , \psib_i(0)\rangle$ are parallel to the initial
bra $\langle \nuf,\rhof|$ and ket $|\nui, \rhoi\rangle$ respectively. 
To bypass this constraint, we  introduce a singular term $h_{i,0} 
\delta(t-0) + h_{i,T} \delta(t-T)$ in the source $h_i(t)$ --- it is 
sufficient to modify $h_i(t)$ only and let $g_i(t)$ be regular, and we 
have also verified that singularities in $h_i(t)$ and $g_i(t)$ at 
times $t \in ]0,T[$ are absent unless $\phib_i(t)$ or $\psib_i(t)$ are
discontinuous, and that a discontinuity of the order parameters
is not favorable in terms of action and can be discarded. 
Optimization of $\Pi_{0,0}$ with respect to $h_{i,0}$ and $h_{i,T}$ yields
\begin{equation}
h_{i,0} = \psib_i (0) + \ln \left[ \frac{\phib_i(0)}{1-\phib_i(0)} \right] 
 -\nui \quad ,
\qquad h_{i,T} = \nuf - \psib_i(T) \quad ,
\end{equation}  
and allows to fulfill the constraints (\ref{constraint_field}) at initial
and final times.

Gathering all contributions to $\Pi_{0,0}$ and using Stirling's formula, 
we find after some algebra
\begin{eqnarray}
\label{Pi_00}
\Pi_{0,0} &=& \sum_i \left[
\int_0^T \dd t\; \psib_i(t) \ddt{\phib_i(t)} +
 \nuf (\rhof - \phib_i(T)) - \nui (\rhoi - \phib_i(0))
 - \phib_i(0) \; \ln\phib_i(0) - (1-\phib_i(0))\; \ln( 1-\phib_i(0) ) \right] 
 \nonumber \\
&+& N \big[ \rhoi \;\ln \rhoi + (1-\rhoi)\; \ln(1-\rhoi) \big]
\end{eqnarray}
Notice that the sum of the last two terms in $\phib_i(0)$ in \eqn 
(\ref{Pi_00}) is equal to the entropy of $N$ noninteracting particles at 
density $\phib_i(0)$.

\subsubsection{Perturbative expansions in powers of $\lambda$ and $\mu$.}
 
For the rest of this section, we call average of an operator $\hat{A}$,
and denote by $\langle \hat{A} \rangle$, the ratio $\langle \nuf,
\rhof|\hat{A} \; \hat{\cal W} (T,0)  | \nui, \rhoi \rangle$ over
$\langle \nuf, \rhof | \hat{\cal W} (T,0)  | \nui, \rhoi \rangle$.  
Let us introduce the operators $\hat{S}_1 = -\int_0^T \dd t\;  
\hat{W}_\mathrm{ann}$ and $\hat{S}_2 = -\int_0^T \dd t\;  
\hat{W}_\mathrm{cre}$. These are directly related to $\Pi$ through the
relations $\partial_\mu \Pi = \langle -\hat{S}_1 \rangle$ and
$\partial_\lambda \Pi = \langle -\hat{S}_2 \rangle$, valid for any
$\lambda$ and $\mu$. The average values of the operators $\hat{S}_1$ and
$\hat{S}_2$ are $\langle \hat{S}_1 \rangle = - \int_0^T \dd t\; \sum_i
\phib_i(t) (\exp(\psib_i(t))-1)$ (for all $\lambda$, $\mu$) and $\langle
\hat{S}_2 \rangle_{\lambda=0}= - \int_0^T \dd t \sum_i \phib_i(t) \sum_{j
\in i} \chib_j(t)/z$ (only for $\lambda=0$)  respectively, and give the
beginning of the expansion of $\Pi$. The sequel is obtained by iterative
application of the following identities true for any (differentiable)
operator $\hat{A}$,
\begin{equation}
\label{identites_dmu_dlambda}
 \partial_\mu \langle \hat{A} \rangle = \langle
  \partial_\mu \hat{A} \rangle + \langle \hat{A}\; \hat{U} \rangle, \quad
 \partial_\lambda \langle \hat{A} \rangle = \langle \partial_\lambda
  \hat{A} \rangle + \langle \hat{A}\; \hat{V} \rangle
\end{equation}
with 
\begin{eqnarray}
\label{definition_U}
\hat{U} & = & -\hat{S}_1 + \langle \hat{S}_1 \rangle + \int_0^T \dd t
\sum_i \Big[ \partial_\mu h_i(t) \;\big( \hat{\phi}_i - \phib_i(t) \big) +
\partial_\mu g_i(t) ( \hat{\chi}_i - \chib_i(t) ) \Big] \\
\label{definition_V}
\hat{V} & = & -\hat{S}_2 + \langle \hat{S}_2 \rangle + \int_0^T \dd t
 \sum_i \Big[ \partial_\lambda h_i(t)\;
     \big( \hat{\phi}_i - \phib_i(t)\big) +
   \partial_\lambda g_i(t)\; \big( \hat{\chi}_i - \chib_i(t) \big) \Big]
\end{eqnarray}
Though sites are coupled by $\hat{V}$ through $\hat{S}_2$, successive
applications of operators $\hat U, \hat V$ permit to evaluate the
(derivatives of) CFs required to compute the expansion in powers of $\mu$
and $\lambda$ at $\lambda=0$ where these CFs are factorized over sites.
Notice that $\hat{U}$, $\hat{V}$ and their derivatives with respect to
$\mu$ and $\lambda$ that will appear in higher orders of perturbation
involve the derivatives of the fields $h$ and $g$. These can be expressed
in a convenient way from the identities $h_i(t) = - \partial_{\phib_i(t)}
\Pi$, $g_i(t) = - \partial_{\chib_i(t)} \Pi$, $h_{i,T} =-
\partial_{\phib_i(T)} \Pi$ and $h_{i,0} = -\partial_{\phib_i(0)} \Pi$. In 
particular, $\partial_\lambda h_{i,T} = \partial_{\phib_i(T)} \langle 
\hat{S}_2 \rangle = 0$ and $\partial_\lambda h_{i,0} = 
\partial_{\phib_i(0)} \langle \hat{S}_2 \rangle = 0$, allowing us to 
remove all terms involving $\delta$'s in $h_i$ in the expressions of
$\hat{U}$ and $\hat{V}$. Also, $\partial_\lambda h_i(t) =
\partial_{\phib_i(t)} \langle \hat{S}_2 \rangle$ and $\partial_\lambda
g_i(t) = \partial_{\chib_i(t)} \langle \hat{S}_2 \rangle$, so that the
operator $\hat V$ writes in a convenient form when $\lambda$ vanishes:  
$\hat{V}_{\lambda=0} = \int_0^T \dd t \sum_i \sum_{j \in i} \phit_i(t)
\chit_j(t) /z$, where $\phit_i(t) = \hat \phi_i - \phib_i(t)\, \id$ and
$\chit_i(t) = \hat{\chi}_i - \chib_i(t)\, \id$ express the deviations of
the elementary operators with respect to their average values: $\langle
\phit_i(t)  \rangle = \langle \chit_i(t) \rangle = 0$ for all $\lambda,
\mu$. Finally, let us give some useful properties of $\hat{U}$ and
$\hat{V}$:
\begin{enumerate}
\item $\langle \hat{U} \rangle = \langle \hat{V} \rangle = 0$
(and if we write these operators in a natural way as sums over the graph
sites, each summand $\langle \hat{U}_i \rangle$ and $\langle \hat{V}_i
\rangle$ also vanishes);
\item  $\langle \hat{U} \phit_i(t) \rangle = \langle
\hat{V} \phit_i(t) \rangle = \langle \hat{U} \chit_i(t) \rangle = \langle
\hat{V} \chit_i(t) \rangle = 0$;
\item Any CF of the type $\langle \hat{A}_1 \, \hat{A}_2  \ldots
 \hat{A}_k  \, \hat{U} \rangle$ vanishes if the time attached to
the $\hat U$ operator is smaller or larger than all other times attached
to the $A_\ell$ operators. Indeed, a direct evaluation of $\langle \nuf, 
\rhof | \hat{U}$ and $\hat{U} | \nui, \rhoi \rangle$ after explicitly 
expressing $\hat{U}$ in a similar way as we did for $\hat{V}_{\lambda=0}$ 
shows that both vectors vanish.
\end{enumerate}
We now apply the above scheme to the calculation of $\Pi$.

\section{Mean-field theory ($z \to \infty $ limit)}

In this Section, we first expose the analysis of the Contact
Process on a complete graph.
We then explain how this mean-field theory can be found back through the 
formalism developed in Section \ref{secfieldtheory}. 

\subsection{A simple derivation of mean-field theory: the Contact Process 
on the complete graph}

Consider CP on the complete graph with $N$ sites. 
As any two sites are adjacent, an exact account of the dynamics can
be obtained from tracking the 
probability $p_n(t)$ that $n$ sites are occupied at time
$t$ \cite{dickman-proccont-metastable-champ-moyen}. The master equation
for these probabilities reads,
\begin{equation} \label{master}
\ddt{p_n}(t) = (n+1) \;  p_{n+1}(t) + \frac{\lambda}{N-1} \; (n-1) \;
(N-n+1)\; p_{n-1}(t)- \bigg( n + \frac{\lambda}{N-1}\;n\; 
(N-n) \bigg) \; p_n (t)  
\quad ,
\end{equation}
with the conventions  $p_{-1}(t)=p_{N+1}(t)=0$. In the large $N$ limit,
we expect from the considerations of Section \ref{secpheno} the following 
scaling behaviour for the probabilities~\cite{langer-metastable}, 
\begin{equation} \label{scalhypo}
p_n(t) = \exp \big(N \,\picm(n/N,t) \big) \quad .
\end{equation}
Inserting this scaling Ansatz into the master equation (\ref{master}) 
yields the equation of motion for the large deviation function $\pi_{MF}
(\rho , t)$ of the density $\rho$ of particles,
\begin{equation}
 \label{equation_evolution_pi_de_rho}
\partial_t \picm(\rho,t) = \tilde{W}_\mathrm{MF} \big[ \rho, 
\partial_\rho  \picm (\rho, t) \big]
\end{equation}
where
\begin{equation}\label{wmfdef0}
\tilde{W}_\mathrm{MF}\big[ \phi,\psi \big] := \phi\; \big(\exp(\psi) - 1\big) 
+ \lambda\; \phi \;(1-\phi) \;\big( \exp(-\psi)-1 \big)
\quad .
\end{equation} 
The very strong analogy between the above definition and the 
expression for the matrix elements of the creation and annihilation 
operators in \eqn (\ref{entriescreaanni}) will be explained
in next Section.

After a transient depending on the initial condition e.g. all sites
are initially occupied and all densities but $\rho=1$ have zero probability, 
the large deviation function $\picm$ relaxes to its stationary value,
\begin{equation}
 \label{pi_de_rho_quasi_stationnaire}
\picm^* (\rho) = -\frac 1{\lambda} + (1-\rho)\; \big(1-\ln \lambda-
\ln(1-\rho)\big)
\quad .
\end{equation}
It is maximal and equal to zero in $\rho=\rho^*$. The value in $\rho =0$
gives immediate access to the average lifetime of the metastable state 
with density $\rho ^*$, that is, the time it takes to the system to reach
the empty state, see \eqn (\ref{ret}), 
\begin{equation}
t_{vac} (\lambda , N) \sim \exp \bigg[ N \; \bigg( 
\frac 1\lambda + \ln \lambda -1 \bigg) \bigg]  \quad .
\end{equation}
This value may be successfully compared to the results of Fig.~4 in
Ref.~\cite{dickman-proccont-metastable-champ-moyen}. A direct numerical
simulation of the evolution of the $p_n(t)$ with $N$ up to $\approx 100$
has allowed us to check the validity of scaling hypothesis
(\ref{scalhypo}) and to obtain a perfect agreement of the experimental
distribution with \eqn (\ref{pi_de_rho_quasi_stationnaire}).

\subsection{Mean-field theory from the ``quantum'' formalism}

The first term, $\Pi_{0,0}$, in the expansion of the effective potential $\Pi$
in powers of $\lambda$ and $\mu$ is given by \eqn (\ref{Pi_00}).
Then $\Pi_{1,0} = -\langle \hat{S}_1 \rangle_{0,0}$
and $\Pi_{0,1} = -\langle  \hat{S}_2 \rangle_{0,0}$, the
expressions of which in terms of $\phib$ and $\chib$ are given above.
These are sufficient to establish the
expressions of $\Pi_0 (\mu)$ and $\Pi_1 (\mu)$, due to the
third property of Section II.B.3. Indeed,
\begin{itemize}
\item $\Pi_{a,0} = 0$ for all $a \ge 2$. \emph{Proof:} 
$\partial_\mu \Pi = -  \langle \hat{S}_1\,
\hat{U} \rangle $ since $\partial_\mu \hat{S}_1 = 0$. 
The expression of $\langle \hat{S}_1 \hat{U} \rangle$ involves a double sum 
over the sites, say, $i,j$, of terms of the form 
$\langle \hat{A}_i \, \hat{U}_j \rangle$. Any such
CF vanishes for all $\mu$ as stated in Section II.B.3.
\item $\Pi_{a,1} = 0$ for all $a \ge 1$. \emph{Proof:}
$\partial_\mu \partial_\lambda \Pi
= -\langle \hat{S}_2 \,\hat{U} \rangle$ which, for $\lambda=0$ and all
$\mu$, reduces to a triple sum over the sites, say, $i$, $j \in i$
and $k$, of CF of the type $\langle \hat{A}_i \,\hat{B}_j
\, \hat{U}_k \rangle$ where the times attached to $\hat A_i$ and 
$\hat B_j$ coincide. Again, these CF vanish due to the presence of the
$\hat U$ operator. 
\end{itemize}
As a result, setting $\mu=1$, we obtain 
\begin{eqnarray}
 \label{Pi_champ_moyen}
\Pi \big[\rhof,T ; \{\phi ,\chi\} | \rhoi,0\big]
&=& \sum_i \Bigg[ \nuf \big (\rhof - \phib_i(T)\big) - 
\nui \big(\rhoi -  \phib_i(0) \big) \nonumber \\
&-& \phib_i(0) \ln(\phib_i(0)) - (1-\phib_i(0)) \ln(1-\phib_i(0))
 + \rhoi \ln \rhoi + (1-\rhoi) \ln(1-\rhoi) \nonumber \\
&+& \int_0^T \dd t \left( \psib_i(t) \ddt{\phib_i}(t)  +
 \phib_i(t) (\exp(\psib_i(t)) - 1) +  \frac{\lambda}{z}\; \phib_i(t)
 \sum_{j \in i} \chib_j(t) \right) \Bigg] + O(\lambda ^2 )
\end{eqnarray}
For $\lambda=0$ resp. for $\mu=0$, we are able to compute the
action~(\ref{Pi_champ_moyen}) directly, without using such a perturbative
expansion as~(\ref{double_developpement_Pi}), since the matrix of
$\hat{W'}$ is upper resp. lower triangular in the basis $(|0\rangle,
|1\rangle)$. However, when both $\lambda$ and $\mu$ are nonzero, we do not
know how to proceed without the perturbative expansion.

In search for a translationally invariant {\em i.e.} site-independent 
evolution, we choose all quantities to be site independent 
and remove site indices. We also drop the bars over the fields
to simplify notations. Equation (\ref{Pi_champ_moyen}) then becomes
$\picm = \Pi/N$ with
\begin{eqnarray}
 \label{Pi_champ_moyen_symetrique}
\picm \big[\rhof,T ; \{\phi ,\chi\} | \rhoi,0\big] 
& = & - \phi(0) \ln(\phi(0)) - (1-\phi(0)) \ln(1-\phi(0)) +
 \rhoi \ln \rhoi + (1-\rhoi) \ln(1-\rhoi)  \nonumber \\
&+& \nuf (\rhof - \phi(T)) - \nui (\rhoi - \phi(0)) +
 \int_0^T \dd t \left( \psi(t) \ddt{\phi}(t) + 
 \tilde{W}_\mathrm{MF}\big[ \phi(t),\psi(t)\big] \right)
\end{eqnarray}
where $\tilde{W}_\mathrm{MF}$ is defined in \eqn (\ref{wmfdef0}). 
The above expression suffices to the study the mean-field (MF) case
{\em i.e.} when the site connectivity $z$ goes to infinity. This may 
happen for large complete graphs, where each site is connected to all 
other sites ($z=N-1$), or in the $D \to \infty$ limit of a $D$-dimensional 
regular lattice ($z=2D$).
As shown in Section IV.A, higher order terms in the $\lambda$ expansion
give $O(1/z)$ additive contributions to $\Pi$ (within a site independent
Ansatz), and vanish in the $z \to \infty$ limit. Therefore, $\picm$ is the
exact mean-field expression for the action.
To obtain the (exponentially in $N$) dominant trajectory of the 
order parameters  $\phi(t),\psi(t)$, we first extremize 
(\ref{Pi_champ_moyen_symetrique}) with respect to $\nuf,\nui$,
to get the expected relations $\phi(T)=\rhof$ and $\phi(0) = \rhoi$.
Notice that, if the initial state were a superposition of configurations 
with various densities e.g.
$\int_0^1 \dd \rhoi \exp(N q(\rhoi)) |\rhoi\rangle$, \eqn 
(\ref{Pi_champ_moyen_symetrique}) would include an additive 
contribution $q(\rhoi)$; the most probable initial density, $\rhoi$, 
would then be given by the solution of $\partial_\rho q(\rhoi) = \psi(0)$.
The resulting expression, 
\begin{equation} \label{lag}
\picm = \int_0^T \dd t \Big\{ \psi(t)  \;
\ddt{\phi}(t) + \tilde{W}_\mathrm{MF} \big[\phi(t),\psi(t) \big] 
\Big\} \quad ,
\end{equation}
has then to be functionally extremized over the fields $\phi(t)$ and 
$\psi(t)$. The equations of motion (EM) for the fields are the 
Hamilton-Jacobi equations associated to Lagrangian (\ref{lag}), 
\begin{equation}
 \label{equations_du_mouvement}
 \ddt{\psi} (t) = \partial_\phi \tilde{W}_\mathrm{MF} \big[
\phi(t),\psi(t) \big]\quad  ,
\qquad  \ddt{\phi}(t) = - \partial_\psi \tilde{W}_\mathrm{MF} 
\big[\phi(t), 
 \psi(t)\big]
\end{equation}
along with the already established initial and final conditions: $\phi(T)  
= \rhof$ and $\phi(0) = \rhoi$. 

\begin{center}
\begin{figure}
A \epsfig{file=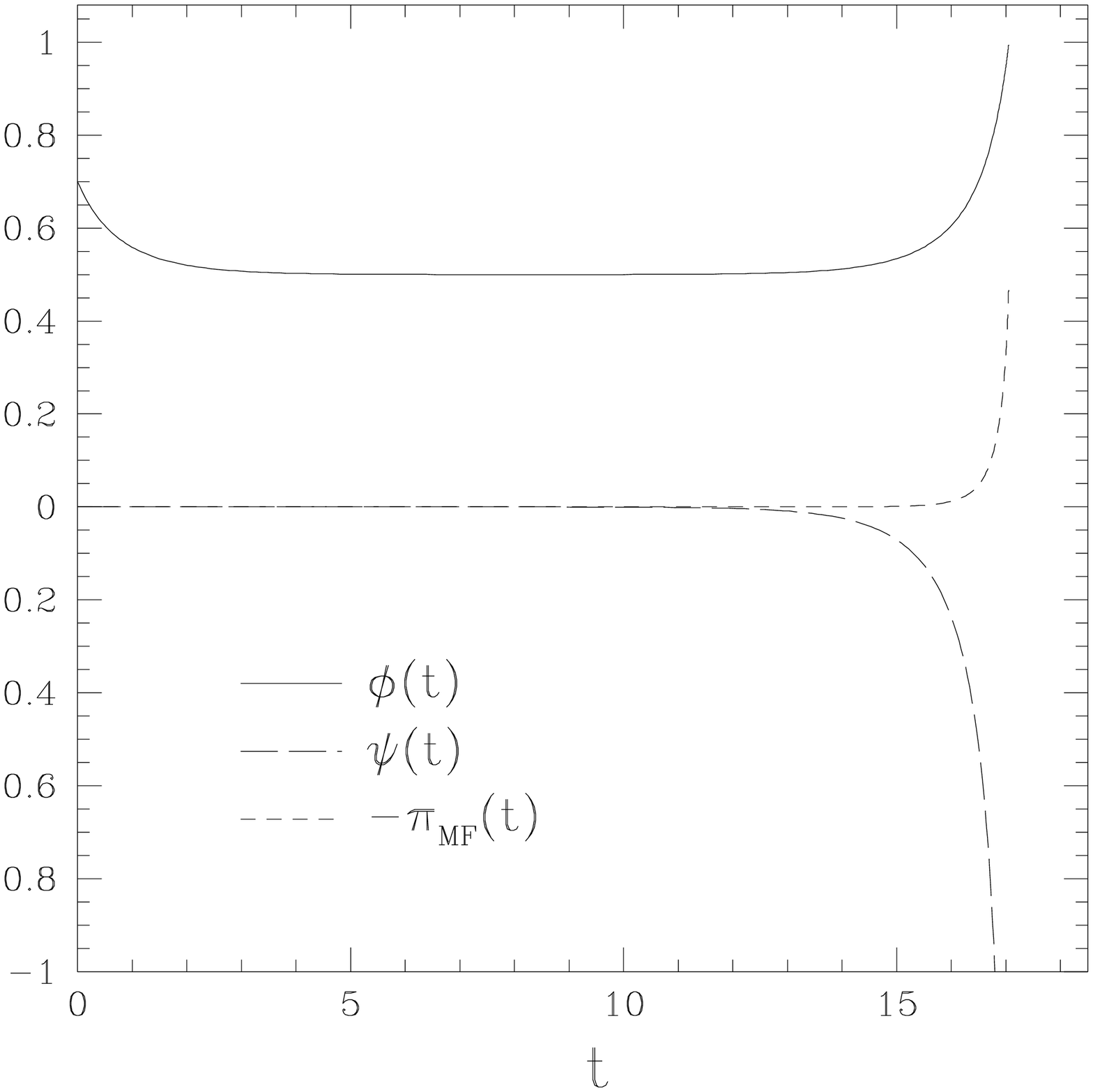,width=6cm}
\hspace{1cm}
B \epsfig{file=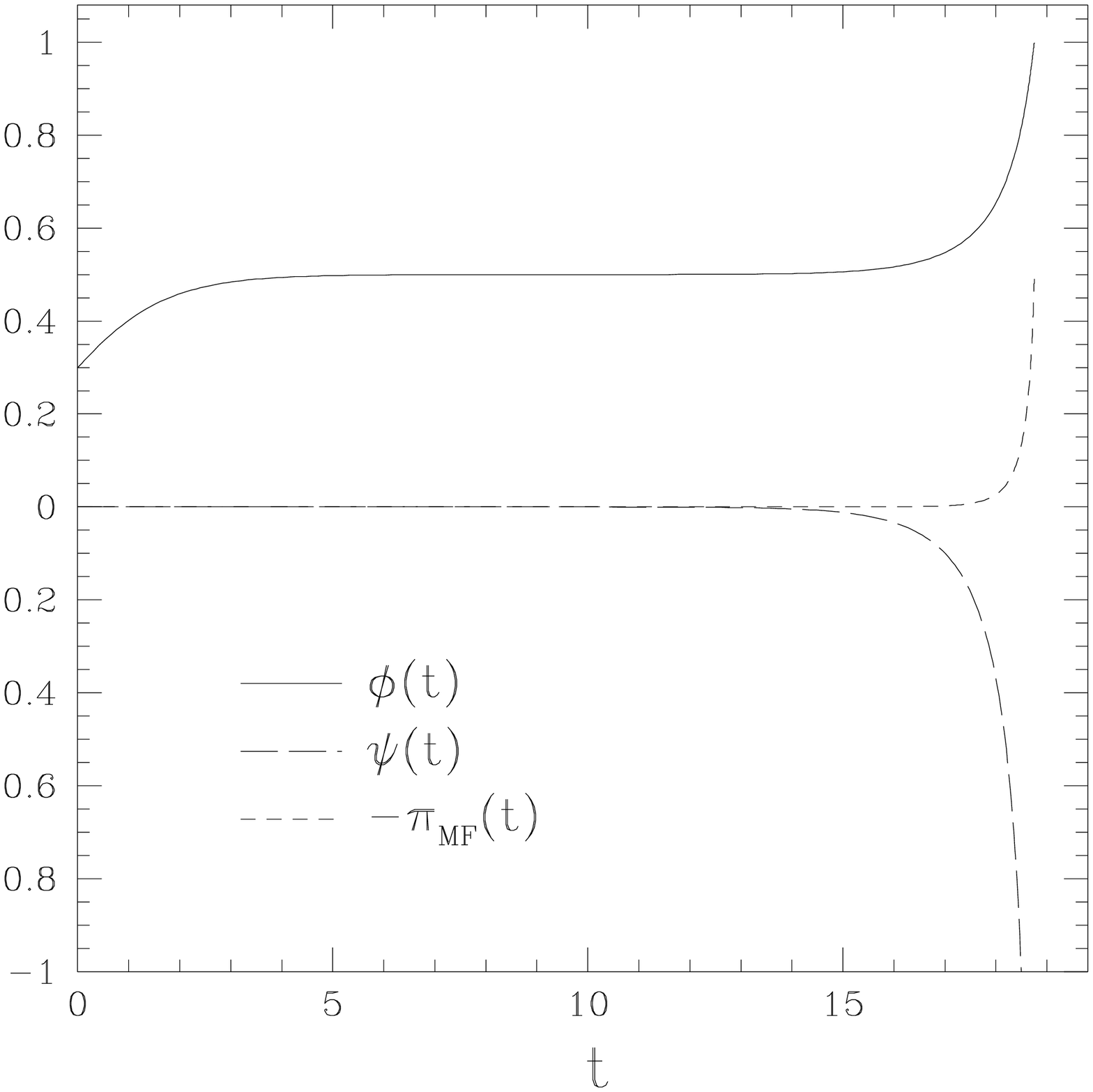,width=6cm}\\
C \epsfig{file=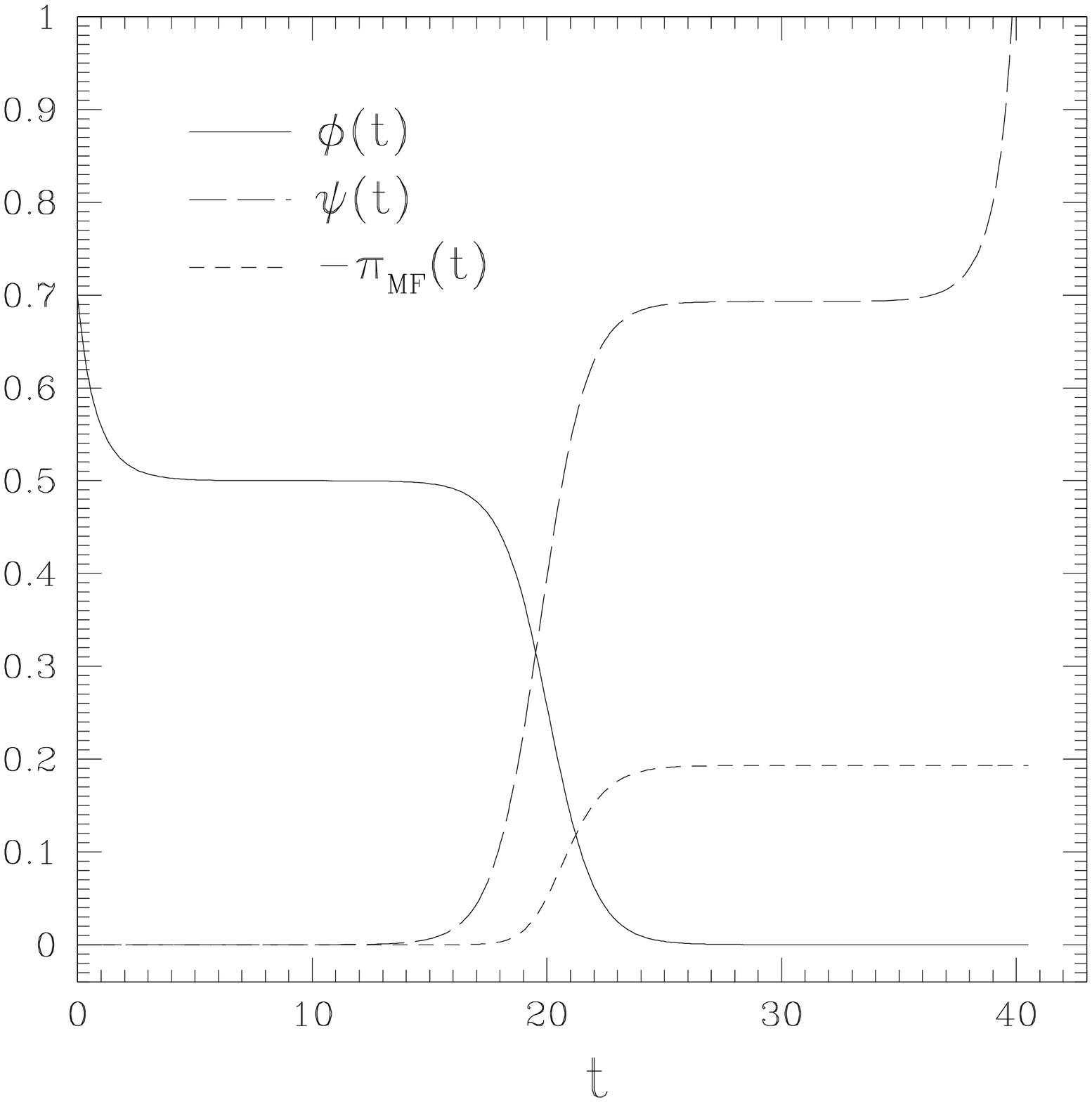,width=6cm}   
\hspace{1cm}
D \epsfig{file=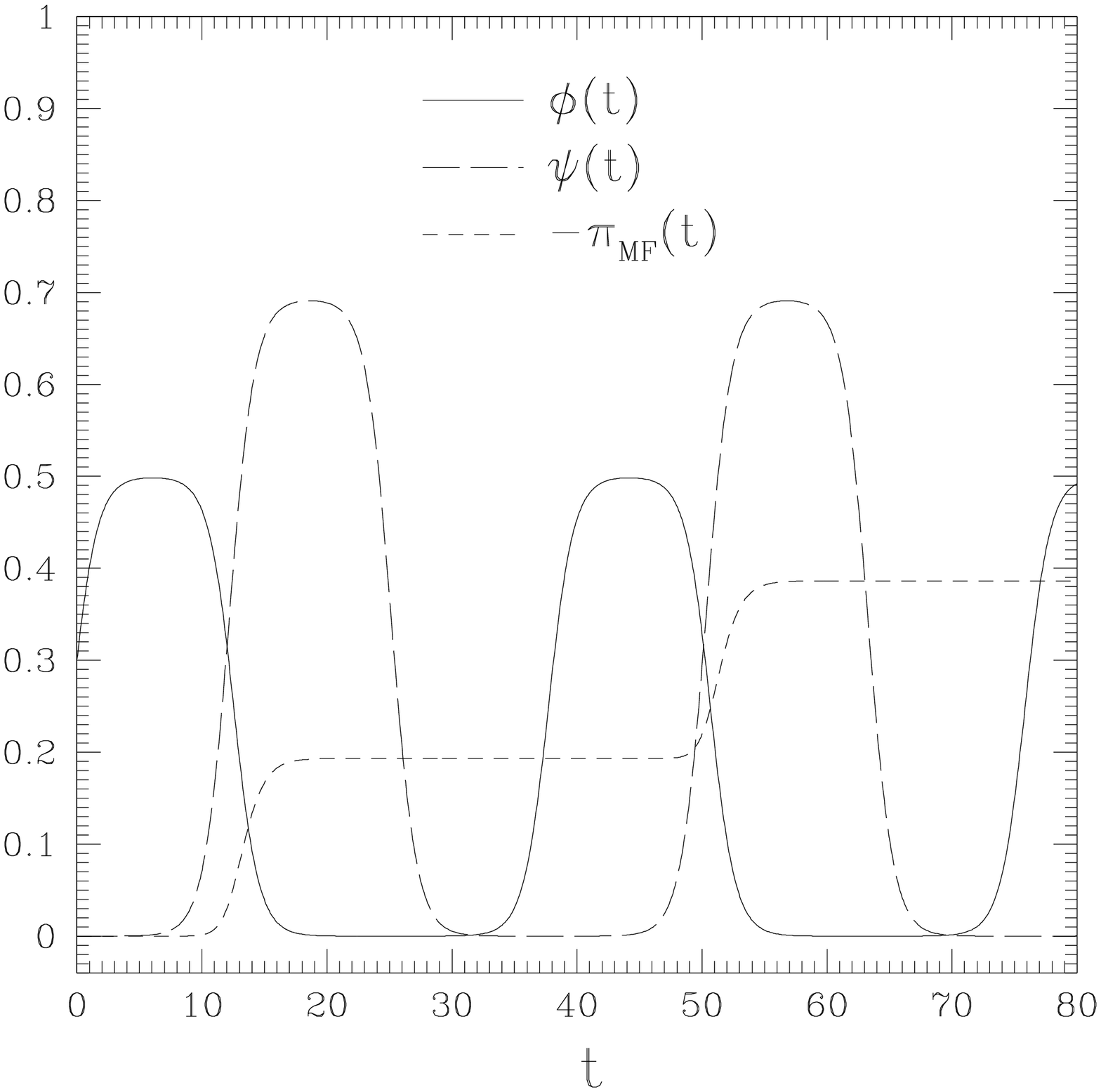,width=6cm}\\
E 
\epsfig{file=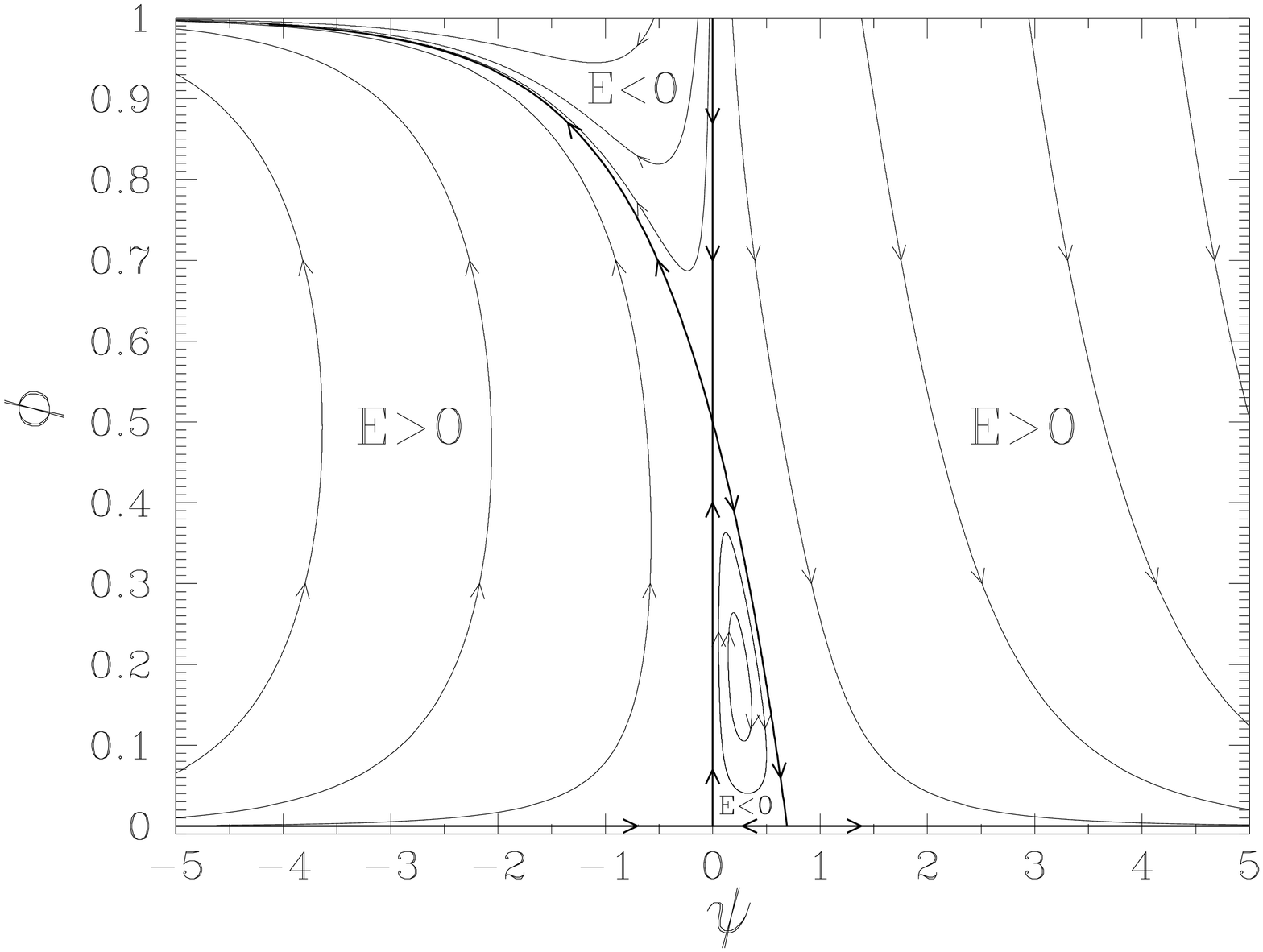,angle=-90,width=11cm,origin=br}
\caption{Solutions of the mean-field equations of
motion~(\ref{equations_du_mouvement}) for $\lambda=2$ ($\rho^*=1/2$).
\textbf{A \& B.} For negative $\psi(0)$ ($= -10^{-7}$ here), the
solutions have a finite lifetime. After quickly reaching the most probable
value of the density of particles, $\rho^*$, $\phi(t)$ stays for a long  
time in the neighborhood of $\rho^*$ but finally  reaches unity,
while $\psi(t) \to -\infty$. The action $\picm$ tends to a negative finite
value; the main contribution to it comes from the final jump of $\phi$ 
from $\rho^*$ to 1.
\textbf{C.} For positive $\psi(0)$ ($=10^{-8}$ here) and $E>0$, 
the lifetime is finite. After quickly reaching $\rho^*$, $\phi(t)$ stays
for a long time in its neighborhood but finally transits to the 
neighborhood of 0 where it stays again for a long time, while $\psi(t)$
transits to $\ln \lambda$ and stays close to it. Finally $\psi(t) \to
+\infty$ while $\phi(t) \to 0$. $\picm$ tends to a negative finite value,
the main contribution of which is accumulated during the transit.
\textbf{D.} For positive $\psi(0)>0$ ($= 10^{-8}$ here) and $E<0$,   
the solution is periodic. $\phi(t)$ and $\psi(t)$ oscillate between two
values inside the $[0,\rho^*]$ and $[0,\ln(\lambda)]$ intervals,
respectively, and the action diverges to $-\infty$ by hops.
\textbf{E.} Phase portrait with all types of solutions (oriented according 
to the time evolution). The phase space divides into four regions 
delimited by the solutions for which $E=0$ (represented in bold lines); in 
each region the sign of $E$ is indicated. The quasistationary state 
appears on this diagram as the crossing point $(\psi=0, \phi=\rho^*)$ of
two solutions with $E=0$: it is stable along the $\phi$ direction but 
instable along the $\psi$ direction. }
\label{sol_EM_typiques}
\end{figure}
\end{center}

The solution of the EM yields the logarithm $N \picm$ of the probability
to go from a configuration with density $\phi(0)$ at time 0 to another
configuration with density $\phi(T)$ at time $T$. To solve these
equations, we make use of the fact that $\tilde{W}_\mathrm{MF}$ is a
conserved quantity from \eqn~(\ref{equations_du_mouvement}), which we
denote by $E$. Then, the density $\phi$ and the time $t$ can be expressed
as functions of $y:=\exp(\psi)$,
\begin{equation}
 \label{t_de_y_champ_moyen}
t = \int_{y (0)}^{y} \dd y' \big[ (y'-\lambda)^2 (y'-1)^2 + 4\, E\, \lambda
\, y' \,  (y'-1) \big] ^{-1/2} \quad .
\end{equation}
The action of the trajectory equals the large deviation function, 
$\picm(\rho,T) = \int_{y(0)}^{y(T)} \dd y'\, \ln(y') \partial_y \phi(y') 
+ T\, E$.
The shape of the solutions of (\ref{equations_du_mouvement}) depends on
the sign of $\psi(0)$. As shown in
Fig.~\ref{sol_EM_typiques}, if we exclude solutions where $\phi(t)$ is
always zero~\footnote{If $\phi(t)$ is always zero, $\psi(t)$ goes to 0 in 
a infinite time if $\psi(0)<\ln(\lambda)$, remains constant if $\psi(0)=0$ 
or $\psi(0)=\ln(\lambda)$, and goes to $+\infty$ in a finite time (all the
shorter as $\psi(0)$ is large) if $\psi(0)>\ln(\lambda)$. All these 
solutions have a vanishing action $\picm$.}, three cases have to be 
distinguished:
\begin{itemize}
\item if $\psi(0)=0$, the solutions have an infinite lifetime. $\psi(t)$ 
remains zero at all times~\footnote{The existence of the solution 
$\psi(t)=0$ to the EM is true beyond mean field, and is a consequence of 
the fact that the evolution operator $\hat W$ conserves the probability.
The latter translates into $(\langle 0 | + \langle 1|) \hat W=0$ and, from
\eqn (\ref{coh}), into $\tilde W(\vec \phi, \vec \psi=0)=0$ for any $\vec
\phi$.}, and the field $\phi(t)$ obeys the first-order ordinary 
differential equation
\begin{equation} \label{dyn56}
\ddt{\phi}(t) = - \lambda \; \phi (t) \; \left( \phi(t) - 1 +
 \frac 1\lambda \right) \quad .
\end{equation}
This equation coincides with the mean-field equation for the
density $\rho(t)$, straightforwardly obtained when neglecting correlations
between occupation numbers of neighboring sites.
At large times, $\phi(t)$ tends to $0$ or $\rho ^* = 1-1/\lambda$ depending
on whether $\lambda$ is smaller or larger than the critical
value $\lambdac=1$ as shown in Fig.~\ref{figures-densite}A. Notice that
the action $\picm$ vanishes as $\psi=0$. 
\item if $\psi(0)<0$, the solutions have a finite lifetime and
$\phi(t) \to 1$ while $\psi(t) \to -\infty$ (see
Fig.~\ref{sol_EM_typiques}A\&B).
\item if $\psi(0)>0$, the sign of the conserved quantity $E$ matters. If
$E > 0$ (which is necessary if $\lambda<1$), the lifetime is finite (see
Fig.~\ref{sol_EM_typiques}C). If $E < 0$, the solutions are periodic (see
Fig.~\ref{sol_EM_typiques}D). If $E=0$, the situation is the natural
intermediate between the two cases. In all cases, the lifetime (or the
period) $T(E)$ increases as $|E|$ diminishes.
\end{itemize}

Nonzero fields $\psi$ select instantonic solutions allowing the system to
escape the typical behaviour described by the $\psi=0$ solution. These
solutions have an extensive action, and are therefore exponentially
suppressed when the system size increases. However, they are the solutions
giving rise to large deviations of the density for a finite volume
(Fig.~\ref{figures-densite}B). The most probable fluctuations correspond
to long times solutions {\em i.e.} $E \to 0$.  In this case, $T(E)$
diverges like $\log E$, and the $T\, E$ term vanishes. When $\lambda >1$,
we have the simple identity $\phi(y)= 1-y/\lambda$ if $0 \le y \le
\lambda$, $0$ if $y \ge \lambda$.  In particular, the weight of the
instantonic solution that goes from $\phi(0)=\rho^*$ to 
$\phi(+\infty)=\rho$ coincides with the stationary large deviation 
function for the
density, $\picm ^* (\rho)$, defined in (\ref{pi_de_rho_quasi_stationnaire}). 

The large deviation function can be obtained at any finite time $t$ too.
To do so, we use expression (\ref{Pi_champ_moyen_symetrique}) to express
the probability of going from the state with
distribution $\pi(\rho,t)$ at time $t$ to the one the state with 
$\pi(\rho,t+\dd t)$ at time $t+\dd t$. Special care must be paid 
to discretizing the term involving time derivations in a symmetric way 
between $t$ and $t+\dd t$ as requested for path 
integrals~\cite{schulman-livre}. The resulting evolution 
equation for the large deviation function coincides with 
(\ref{equation_evolution_pi_de_rho}) as expected.

\subsection{Relationship with Martin-Siggia-Rose and Janssen-de 
Dominicis formalisms}

The above formalism is, to some extent, related to the treatment of
Langevin equations by Martin, Siggia and Rose (MSR)
\cite{martin-siggia-rose} and Janssen and de Dominicis
\cite{janssen-intchem-martin-siggia-rose, dedominicis-jphyscolloq-1976,
bausch-janssen-wagner-intchem-martin-siggia-rose}.
If $x(t)$ is a classical (scalar or vector) field, the Langevin equation
\begin{equation}
\ddt{x}(t) = -V'(x(t)) + \eta(t)
\end{equation}
describes its evolution in the potential $V$ ($V'$ denotes $\partial_x V$)  
under the random Gaussian force $\eta$. $\eta$ is specified by its first
two moments:  $\overline{\eta(t)}=0$, $\overline{\eta(t) \eta(t')}=2T
\delta(t-t')$ where $T$ is the temperature and the overlines denote the
noise average. Let $P(x_T,T|x_0,0)$ be the probability that $x$ equals
$x_T$ at time $T$ conditioned to its initial value. This probability can
be expressed as a path integral through the introduction of a response
variable $\hat{x}$ conjugated to $x$~\cite{martin-siggia-rose,
janssen-intchem-martin-siggia-rose, dedominicis-jphyscolloq-1976,
bausch-janssen-wagner-intchem-martin-siggia-rose},
\begin{equation}
 \label{integrale_de_chemins_pour_equation_de_langevin}
P(x_T,T|x_0,0) = \int_{x(0)=x_0} ^{x(T)=x_T}\mathcal{D}x(\tau) \;
\mathcal{D}\hat{x}(\tau) \; \exp \left[ \int_{0}^t d\tau \left( 
i \hat{x} (\tau ) \left( \ddt{x}(\tau) +
   V'(x(\tau)) \right) - T \hat{x}^2(\tau) \right) \right] \ .
\end{equation}
Functional optimization of 
(\ref{integrale_de_chemins_pour_equation_de_langevin}) with 
respect to $x,\hat{x}$ yields the classical equations of motion,
\begin{equation}
  \label{EM_pour_equation_de_langevin}
 \ddt{x(t)} = - V'(x(t)) - i\, 2T \,\hat{x}(t) , \quad
 \ddt{\hat{x}(t)} = \hat{x}(t)\; V''(x(t))
\end{equation}
from which we obtain that ${\ddt{x}(t)}^2-{V'(x(t))}^2$ is a constant of 
motion.

These EM are formally identical to those derived for 
CP~(\ref{equations_du_mouvement}) in Section III ($x$ playing the role 
of $\phi$ and $\hat{x}$ that of $-i\,\psi$) when $|\psi|<<1$, with the
following choice of potential and temperature,
\begin{equation}
V'(\phi) = \lambda \, \phi\, (\phi-\phi^{*}) , \quad
2\, T(\phi) =  \phi\big( 1+\lambda \, (1-\phi) \big)
\end{equation}

Therefore, in the weakly fluctuating regime where $\psi$ is small and
$\phi$ close to its most probable value $\rho^*$, the system evolves in an
effective potential $V$ with fluctuations that can be described by a
Langevin equation at temperature $T$~\footnote{Notice that the condition
$|\psi|\ll 1$ demands that the density $\phi$ stays close to
its equilibrium value $\phi ^*$, and thus $T$ is essentially 
density independent and equal to $T(\phi^*)$.}. This
description breaks down as soon as we consider large deviations, or
transitions from the metastable to the empty state, as can be seen
from the numerical comparison of the true solutions
of \eqn (\ref{equations_du_mouvement}) with the solutions
of \eqn (\ref{EM_pour_equation_de_langevin}).

Extending the validity of the Langevin equation approach to the full
domain of $\phi$ and $\psi$ would require the use of a MSR-like
formalism~\cite{martin-siggia-rose, janssen-intchem-martin-siggia-rose,
dedominicis-jphyscolloq-1976,
bausch-janssen-wagner-intchem-martin-siggia-rose}. Interestingly, MSR
stated in their original work that the knowledge of the physical field 
$\phi$ alone is not sufficient to compute all quantities of interest 
beyond the Gaussian approximation, and that the introduction of a second 
operator (corresponding to our $\psi$, or to $a$ and $a^+$, whereas $\phi$
corresponds to $a^+a$) to express the response functions of the physical
field was required.

\section{Finite-dimension theory ($1/z$ expansion)}

\subsection{Analytical calculation}

Calculation of the corrections to mean-field theory requires the knowledge
of higher-order terms in the expansion of $\Pi$ in powers of $\lambda$ and
$\mu$. Strictly speaking, our expansion is, after we resum the 
$\mu$-expansion, an expansion in powers of $\lambda$, or, more precisely, 
of $\lambda/z$. It naturally gives access to an expansion of $\Pi$ in 
powers of $1/z$ as shown below.

\subsubsection{The diagrammatic expansion}

To show how this expansion works in practice, let us consider the first 
correction to \eqn (\ref{Pi_champ_moyen_symetrique}), 
$2\,\Pi_{2,0}=\partial^2_\lambda 
\Pi|_{0,0}=-\partial_\lambda\langle\hat{S}_2 \rangle_{0,0} = - \langle 
\hat{S}_2 \, \hat{V} \rangle_{0,0} = \langle \hat{V}^2 \rangle_{0,0}$
since $\partial_\lambda \hat{S}_2 = 0$ and $\langle
\hat{S}_2 \hat{V} \rangle = - \langle \hat{V}^2 \rangle$. $\hat{V}_{0,0}$
involves a sum over couples of neighboring sites that can be written as a
sum over oriented links. The term $\phit_i \,\chit_j$ will be represented
by a link going from site $i$ to site $j$. Therefore $\langle \hat{V}^2
\rangle_{0,0}$ writes as a sum over couples of oriented links. When
$\lambda=0$, sites decouple and terms where a site carries a single
operator $\phit$ or $\chit$ vanish (Section II.B.3). We are thus 
left with terms where two parallel links loop over two neighboring 
sites $i,j$ on the lattice,
\begin{equation} \label{para}
\mbox{$i$\ \epsfig{file=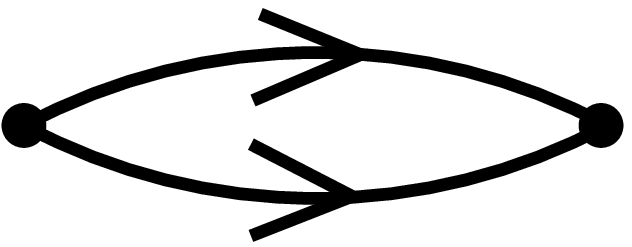,height=8pt}$j$}
= \int_0^T \dd t
\int_0^T \dd t' \;\langle \phit_i(t)\, \chit_j(t)\, \phit_i(t')\, 
 \chit_j(t') \rangle_{0,0} \qquad ,
\end{equation} 
and terms with the same structure but links pointing in opposite
directions. As $\lambda=0$, this four-point CF factorizes into a product
of two two-point CFs:
$ \langle \phit_i(t)\, \chit_j(t)\, \phit_i(t')\, \chit_j(t')
\rangle_{0,0} = \langle \phit_i(t)\, \phit_i(t') \rangle_{0,0} \;
\langle \chit_j(t)\, \chit_j(t') \rangle_{0,0}$. Each of these CFs can be 
straightforwardly computed e.g.
$\langle \phit_i(t)\, \phit_i(t') \rangle_{0,0} =
\phib_i(t_1)\, (1-\phib_i(t_2))$ where $t_1=\min(t,t')$ 
and $t_2= \max(t,t')$. The final result reads
\begin{equation} \label{para2}
\mbox{$i$\ \epsfig{file=terme_lambda2_oriente.eps,height=8pt}$j$}
= -2 \int_0^T \dd t_2 \int_0^{t_2} \dd t_1\; \phib_i(t_1)
\;\big(1-\phib_i(t_2)\big) \; \big(1+\chib_j(t_1)\big) \; \chib_j(t_2) \quad .
\end{equation}
The antiparallel two-site loop diagram may be evaluated in the same way.
Notice that the final outcome does not depend on the indices $i,j$ of the
sites, provided these are neighbors on the lattice. We will therefore drop
site indices in the following. Then, we count the multiplicity of each
diagram, equal to $N\,z$ for both parallel and antiparallel two-sites
loops. Finally, each diagram gets a factor $1/z^2$ coming from the
$(\lambda/z)^2$ factor in $\hat{W}_\mathrm{cre}$. As a result, the net
contribution will be of the order of $1/z$, that is, $1/D$ on a
$D$-dimensional hypercubic lattice.

We are now able to write the general expression of the $1/z$ expansion in
a graphical way. For instance, for a  hypercubic lattice of dimension $D$ 
(and site connectivity $z=2D$),
\begin{eqnarray} \label{diag_exp}
\pi & = & \picm +
\frac{\lambda^2}{2! (2D)^2} D
 \  \epsfig{file=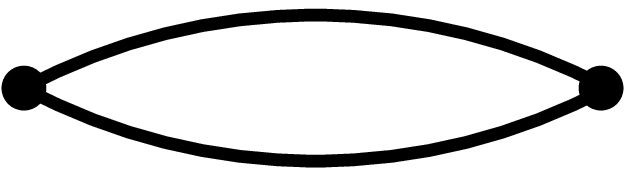,height=6pt} + \nonumber \\ & &
\frac{\lambda^3}{3! (2D)^3} D
 \  \epsfig{file=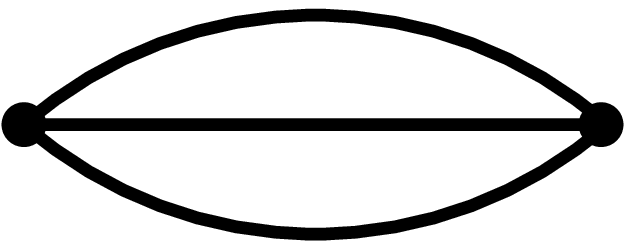,height=8pt} +
\frac{\lambda^4}{4! (2D)^4} D(2D-1) 6
 \  \epsfig{file=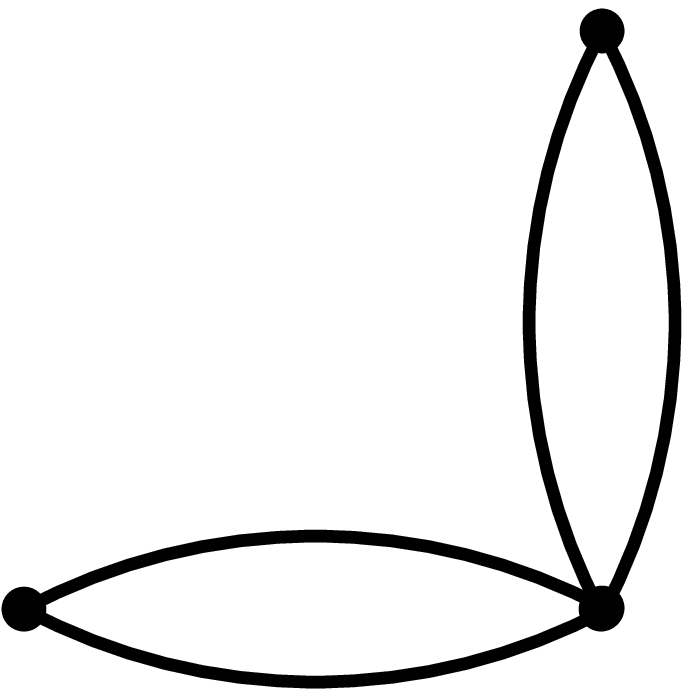,height=18pt} +
\frac{\lambda^4}{4! (2D)^4} \frac{D(D-1)}{2} 4!
 \  \epsfig{file=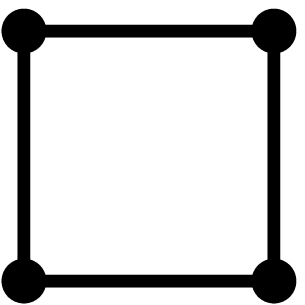,height=10pt} + \nonumber \\ & &
\frac{\lambda^4}{4! (2D)^4} D
 \  \epsfig{file=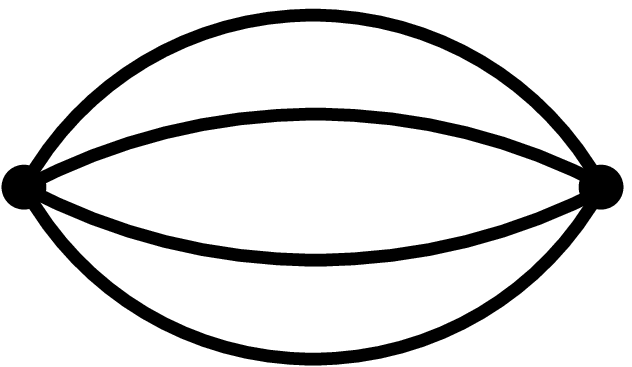,height=12pt} +
\ldots
\end{eqnarray}
Each undirected diagram in the above expansion represents the sum of the
correlation functions (like the one entering $\Pi_{2,0}$ and evaluated in
the previous paragraph) that share the same support on the graph but
differ by the orientations of the links. There are $2^\ell$ link
orientations for an undirected diagram with $\ell$ links. The coefficients
in front of the diagrams take into account the power of $\lambda/z$ (equal
to the number of links in the diagram), the inverse factorial from the
Taylor formula, and a combinatorial factor. This combinatorial coefficient
is the product of the multiplicity of the undirected diagram (number of
times it can be drawn on the lattice, divided by the lattice size $N$) and
of the number of ways to associate a time $t$ to each link of the diagram
when evaluating the CF~\footnote{The situation is in fact a bit more
complex: for some diagrams, one has to compute several families of CFs
(members of a family only differ by the orientations of the links). In
addition to the main family where the average of the product of all link
operators is taken, one has to compute families where link operators are
grouped into separate averages, as if the graph were not connected, and/or
families arising from the derivation with respect to $\phib$ or $\chib$ to
obtain the $\partial_\lambda$'s in the $\hat{V}$ operator. The
combinatorial factor of the undirected diagram is the same for all
families (thus the order in $1/z$ is well defined for the diagram) but the
number of ways to associate times to links may differ.}. As one may infer
from the first diagrams, there is only a finite number of terms
contributing to a given order in $1/z$.

\subsubsection{Summation of the $\mu$-expansion and memory kernels}

$\Pi$ being the logarithm of a generating function of the fields, 
all the diagrams entering expansion (\ref{diag_exp}) are connected. 
Moreover, for $\mu=0$ all nonirreducible diagrams
\emph{i.e.} which may be cut into two, or more, pieces by removal of
a vertex vanish, as in the virial
expansion~\cite{hansen-macdonald} and possibly the Ising
model~\cite{agjy-ising}. This statement does unfortunately
not extend to nonzero $\mu$.
To get the $\mu$-expansion we make repeated uses of the leftmost identity 
in \eqn (\ref{identites_dmu_dlambda}). As $\partial_\mu \hat{U} = 
\partial_\mu \hat{V} =0$, this amounts to insert $\hat{U}$ operators (one 
for each power of $\mu$) in the CFs, yielding contributions of the form
 \begin{equation} \label{def_d}
D_n := \int_0 ^T \dd t_1 \int_0 ^T \dd t_2\ldots \int_0 ^T
\dd t_{n+2}\, \langle \phit_i(t_1)\; \hat{U}_i(t_2)\; \hat{U}_i(t_3) 
\ldots \hat{U}_i(t_{n+1}) \; \phit_i(t_{n+2}) \rangle \quad .
\end{equation}
As stated in Section II.B.3, such a term vanishes if one of the times
$t_2$, $t_3$, \ldots $t_{n+1}$ associated to the $\hat{U}$ operators is 
the minimum or maximum of all times. If not, the integrand is the product 
of factors 
\begin{equation}
\xib(t_k) := - \frac{\exp \psib(t_k)}{1-\phib(t_k)}
\end{equation}
for each $\hat{U} (t_k)$
and a factor depending on the other operators involved in the CF. If
there are strictly less than four other operators in the CF, or no $\hat{U}$
operator between the second and the second last times, this
latter factor is equal to the CF where all $\hat U$ have been removed; 
for instance, the term $D_n$ defined in \eqn (\ref{def_d}) reads
\begin{equation}
D_n = 2\, \int_0 ^T \dd t \int_0 ^t \dd t' \;
\langle \phit_i(t) \; \phit_i(t') \rangle \; \left( 
\int_{t'} ^{t} \dd t'' \; \xib_i (t'') \right) ^n \quad .
\end{equation}
If the original CF involves more operators, or a `badly' located $\hat U$,
there appears a kind of disentanglement of the original operators. 
Consider for instance the four-field CF $D':=
\langle \phit_i(t_1) \phit_i(t_2) \phit_i(t_3) \phit_i(t_4) \rangle$ with
$t_1 < t_2 < t_3 < t_4$. A direct evaluation shows that $D'$ equals
$-\phib_i(t_1) (-1+\phib_i(t_4))
(-\phib_i(t_3) + 1 + 3\phib_i(t_2)\phib_i(t_3) - 2\phib_i(t_2))$ and
cannot be expressed as a product of four factors, each of which would be 
associated to a time $t_j$. We shall say that $D'$ is entangled, here due 
to the $t_2,t_3$ term. Insertion of at least one $\hat{U}_i$ between
$t_2$ and $t_3$, as in $D'' := \langle \phit_i(t_1) \hat{U}_i(t'_1)
\hat{U}_i(t'_2) \ldots \hat{U}_i(t'_{n_1}) \phit_i(t_2)
\hat{U}_i(t'_{n_1+1}) \hat{U}_i(t'_{n_1+2}) \ldots \hat{U}_i(t'_{n_2})
\phit_i(t_3) \hat{U}_i(t'_{n_2+1})  \hat{U}_i(t'_{n_2+2}) \ldots
\hat{U}_i(t'_{n_3}) \phit_i(t_4) \rangle$ with $t_1 < t'_1 < \ldots < 
t'_{n_1} < t_2 < t'_{n_1+1} < \ldots < t'_{n_2} < t_3 < t'_{n_2+1} < 
\ldots < t'_{n_3} < t_4$ removes this entanglement. $D''$ is the product
of factors $\xib_i(t'_k)$, one for each $\hat{U}(t'_k)$, and of the 
disentangled (factorized) expression
$-\phib_i(t_1) (-1+\phib_i(t_4)) (-1+2\phib_i(t_2)) (-1+2\phib_i(t_3))$. 
We conclude that a CF of the type
\begin{equation}
D''_n := \int_0^T \dd t_4 \int_0^T \dd t_3 \int_0^T \dd t_2 \int_0^T \dd 
 t_1 \int_0^T \dd t'_1 \int_0^T \dd t'_2 \ldots \int_0^T \dd t'_n
 \langle \phit_i(t_1) \phit_i(t_2) \phit_i(t_3) \phit_i(t_4) 
  \hat{U}_i(t'_1) \hat{U}_i(t'_2) \ldots \hat{U}_i(t'_n) \rangle
\end{equation}
can be expressed as
\begin{eqnarray}
D''_n & = & 24 \int_0^T \dd t_4 \int_0^{t_4} \dd t_3 \int_0^{t_3} \dd t_2
\int_0^{t_2} \dd t_1 \langle \phit_i(t_1) \phit_i(t_2) \phit_i(t_3) 
\phit_i(t_4) \rangle_\mathrm{dis} \left( \int_{t_1}^{t_4} \dd t \; 
\xib_i(t) \right)^n \nonumber \\
& + & 24 \int_0^T \dd t_4 \int_0^{t_4} \dd t_3 \int_0^{t_3} \dd t_2  
\int_0^{t_2} \dd t_1 \left( \langle \phit_i(t_1) \phit_i(t_2) \phit_i(t_3)
\phit_i(t_4) \rangle - \langle \phit_i(t_1) \phit_i(t_2) \phit_i(t_3)
\phit_i(t_4) \rangle_\mathrm{dis} \right) \times \nonumber \\
& & \quad\quad \left( \int_{t_1}^{t_2} \dd t \; \xib_i(t) + 
\int_{t_3}^{t_4} \dd t \; \xib_i(t) \right)^n
\end{eqnarray}
where the `dis' subscript means that the disentangled expression for
the CF integrand $\langle \ldots \rangle$ must be taken.

It appears that, at any given order in $\lambda$, the series expansion in
powers of $\mu$ can be easily summed up, yielding 
memory kernels in the resulting $\Pi_a (\mu )$ with $a \ge 2$. 
For instance, $\int_0^T \dd t \int_0^{T} \dd t' \;\langle \phit_i(t)  
\phit_i(t') \rangle$ is replaced with
\begin{equation}
\sum_{n \ge 2} \frac{\mu ^n}{ n!} \; D_n 
= \int_0^T \dd t \int_0^{T} \dd t' \; \langle \phit_i(t)
\phit_i(t') \rangle \; \exp \left(\mu \int_{\min (t,t') }^{\max (t,t')} 
\dd t'' \; \xib_i(t'' )\right) 
\quad ,
\end{equation}
the memory kernel being the exponential term, where $\mu$ is eventually 
set to one. Due to the presence of the kernel, the CF between two 
operators at times $t$ and $t'$ decreases with a time difference $|t-t'|$ 
(remember  $\xib < 0$).

This property extends to more than two operators.  The contribution
of a diagram with $n$ links, thus of order $\lambda^n$ in the
$\lambda$-expansion of $\Pi$, may be written as the sum of a finite
number of terms of the form
\begin{equation}
F :=\int_0^T \dd t_n \int_0^{t_n} \dd t_{n-1} \ldots \int_0^{t_2} \dd t_1
 \  P \  \exp\left( i_1 \int_{t_1}^{t_2} \dd t \xib(t) + \ldots
        + i_n \int_{t_{n-1}}^{t_n} \dd t \xib(t) \right)
\end{equation}
where $P$ is a polynomial in $\phi(t_1)$, $\exp(-\psi(t_1))$, \ldots, 
$\phi(t_n)$ and $\exp(-\psi(t_n))$ and $i_1$, $i_2$, \ldots, $i_n$ are 
positive integers. The presence of the memory kernels ensures that the 
integrand sharply decreases as $t_n-t_1$ increases.

One may wonder why memory kernels appear in the expressions above, whereas
CP is a Markovian stochastic process. This is the consequence of the
projection of the complete distribution of states, $|P (t)\rangle$, that
is, the knowledge of the probability of occupation or vacancy of all
sites, onto a partial description where we keep track of the order
parameters $\phi(t),\psi(t)$ only. Beyond mean field, the degrees of
freedom which were discarded pop up as non-Markovian contributions to the
dynamical evolution of the order parameters. This phenomenon is well
known, and can be illustrated with an elementary example proposed in
Appendix A.

\subsection{Results and comparison with simulations}

Following the above recipes, the effective potential $\pi$ may be
expanded in powers of $1/z$, see \eqn (\ref{diag_exp}). 
The explicit expression for the first correction to mean field, coming
from the two-site loop diagram, reads
\begin{equation}
\pi_1 = - \frac{2\lambda^2}z
\int_0^T \dd t_2 \big(1-\phi(t_2) \big)^2 \,\left(e^{-\psi(t_2)}-1\right)
\int_0^{t_2} \dd t_1 \, \phi(t_1)\, \left( \big(1-\phi(t_1)\big) 
e^{-\psi(t_1)} + \phi(t_1) \right) \;
\exp \left(2 \int_{t_1}^{t_2} \dd t'' \xi(t'') \right)
\end{equation}
The $O(1/z^2)$ corrections to $\pi$ required the calculation of the
diagrams listed in \eqn (\ref{diag_exp}). The resulting expression, $\pi
_2$, for the $D$-dimensional hypercubic lattice (where $z=2\,D$) is too
lengthy to be given here, but can be obtained with the help of a computer
algebra software. This gives, as a by-product, the expression $\pi_2$ for
the Cayley lattice (infinite graph without loops where all sites have
exactly $z$ neighbours) after removal of the square diagram
${}_l^i$\epsfig{file=terme_lambda4carre.eps,height=6pt}${}_k^j$.

Functional optimization of the resulting expression for $\pi$ with
respect to $\phi(t)$ and $\psi(t)$ yields EM which include
corrections of orders $1/z$ and $1/z^2$ to the Hamilton-Jacobi equations
(\ref{equations_du_mouvement}). These corrections involve
terms with multiple integrals on the time. 

\subsubsection{Corrections to the density $\rho ^*$ of particles and
the critical parameter $\lambdac$}

We first concentrate on the solution to the EM with vanishing $\psi(t)$ 
at all times $t$. The equation for $\phi (t)$ then simplifies; for instance,
to the first order in $1/z$, we obtain
\begin{equation} \label{dyn57}
\ddt{\phi}(t) = - \lambda \; \phi (t) \; \left( \phi(t) -
1 + \frac 1\lambda \right) - \frac {2\lambda ^2}z \;
\big( 1- \phi (t) \big)^2 \; \int_0 ^t \dd t'\; \phi(t') \;
\exp \left( -2 \int_{t'} ^t \frac{\dd t''}{1-\phi (t'')} \right)  \quad .
\end{equation}
With the help of computer algebra, the EM for $\phi(t)$ can be written
up to order $1/z^2$, and its asymptotic behaviour analyzed. We find 
that $\phi (t\to\infty)$ equals 0 or
$\rho^* >0 $ depending on the value of the parameter $\lambda$ with respect
to its critical value $\lambdac$. The asymptotic density reads 
\begin{equation}
 \label{rhomax_hypercubique}
\rho^* = 1 - \frac{1}{\lambda} - \frac{1}{\lambda^2 z}
         - \frac{6 \lambda^2 + 11 \lambda +3}{6 \lambda^4 z^2}
         + O(1/z^3)
\end{equation}
on the $D$-dimensional hypercubic lattice ($z=2D$), and
\begin{equation}
 \label{rhomax_arbre_cayley}
\rho^* = 1 - \frac{1}{\lambda} - \frac{1}{\lambda^2 z}
         - \frac{6 \lambda^2 + 11 \lambda -3}{6 \lambda^4 z^2}
         + O(1/z^3)
\end{equation}
on the Cayley tree where all sites have $z$ neighbours. The critical value
$\lambdac$ can be easily obtained from the above equations as the value of
the parameter $\lambda$ below which no state with a finite density of
particles can survive. This is intended to be the lower critical value
$\lambda_{C_\mathrm{lower}}$ on the Cayley
tree~\cite{liggett-livre-1999}, where there exists an intermediate range
$[\lambda_{C_\mathrm{lower}},\lambda_{C_\mathrm{upper}}]$ of values of
$\lambda$ where nonuniform metastable states can survive without invading
the whole graph while, above  $\lambda_{C_\mathrm{upper}}$, there is a 
single uniform metastable state --- on hypercubic lattices, both 
thresholds
coincide. Setting $\rho^*=0$, we obtain 
\begin{equation}
 \label{lambdac_hypercubique}
\lambdac = 1 + \frac 1z +  \frac 7{3 z^2} + O(1/z^3)
\end{equation}
on a the hypercubic lattice, and
\begin{equation}
 \label{lambdac_arbre_cayley}
\lambdac = 1 + \frac 1z + \frac 4{3 z^2} + O(1/z^3)
\end{equation}
on the Cayley tree. These results are compatible with the rigorous bounds
$(1-1/(2D))^{-1} \le \lambdac \le 4$ (hypercubic lattice) and $1 \le
\lambdac \le (1-2/z)^{-1}$ (on the Cayley tree)  
\cite{liggett-livre-1985, liggett-livre-1999}. Notice that the lower bound
$\lambdac \ge (1-1/(2D))^{-1}$ was obtained~\cite{liggett-livre-1985}
through a two-site calculation. When discarding all diagrams but the
two-site loops, we find $\lambdac = 1 + 1/(2D) + 1/(2D)^2 + O(1/(2D)^3)$,
that is, the same bound. For small dimensions, our asymptotic results are
of poor quality. We refer the interested reader to the various works that
give more precise estimates for $\lambdac$ e.g. for hypercubic lattices in
$1 \le D \le 5$
ref.~\cite{proccont-nconstant-simulations-sabag-deoliveira}, for $D=1$
through series expansions ref.~\cite{dickman-series-1989,
dickman-series-1994}.

\begin{center}
\begin{figure}
\epsfig{file=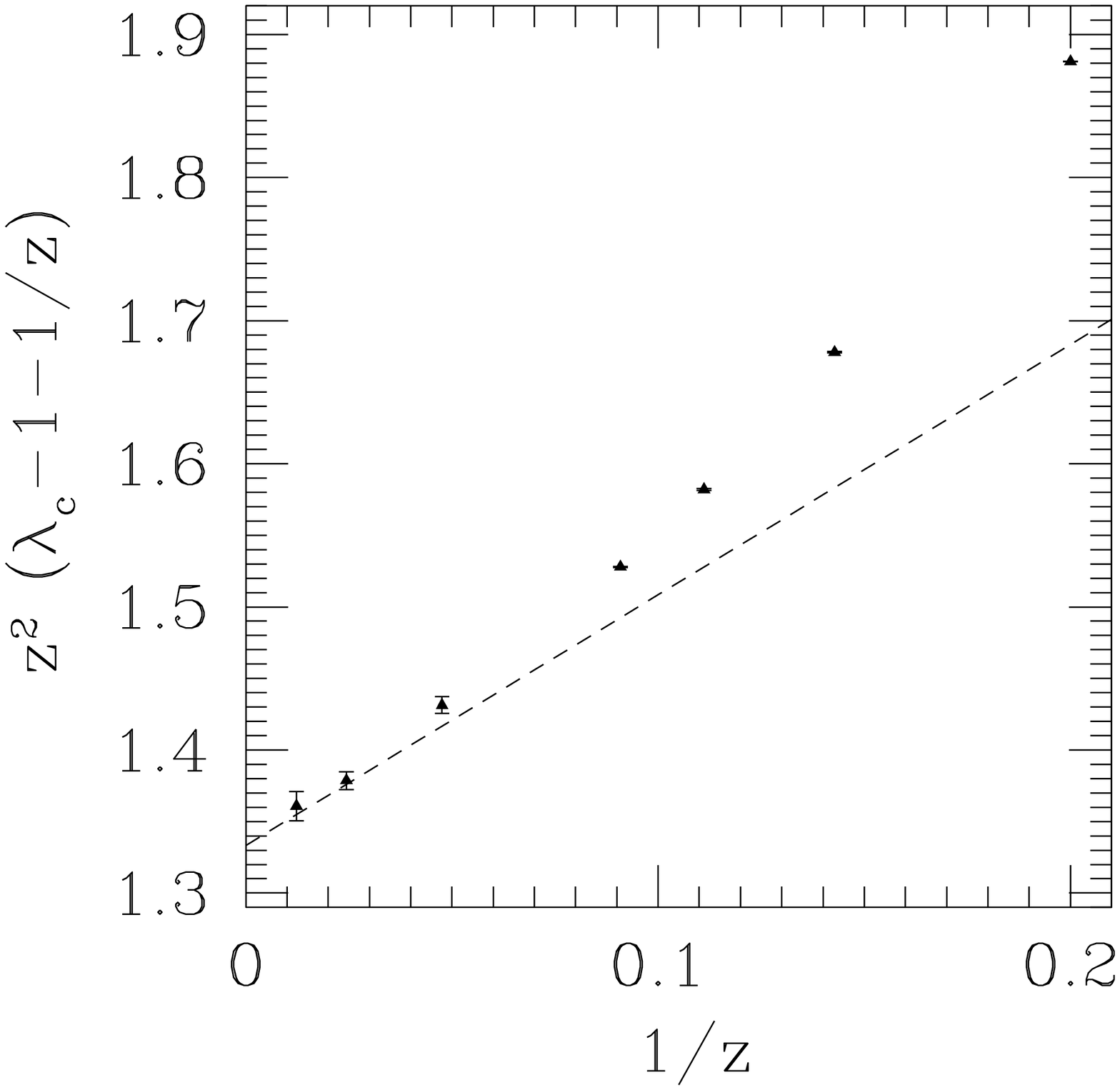,width=7cm}
\hspace{1cm}
\epsfig{file=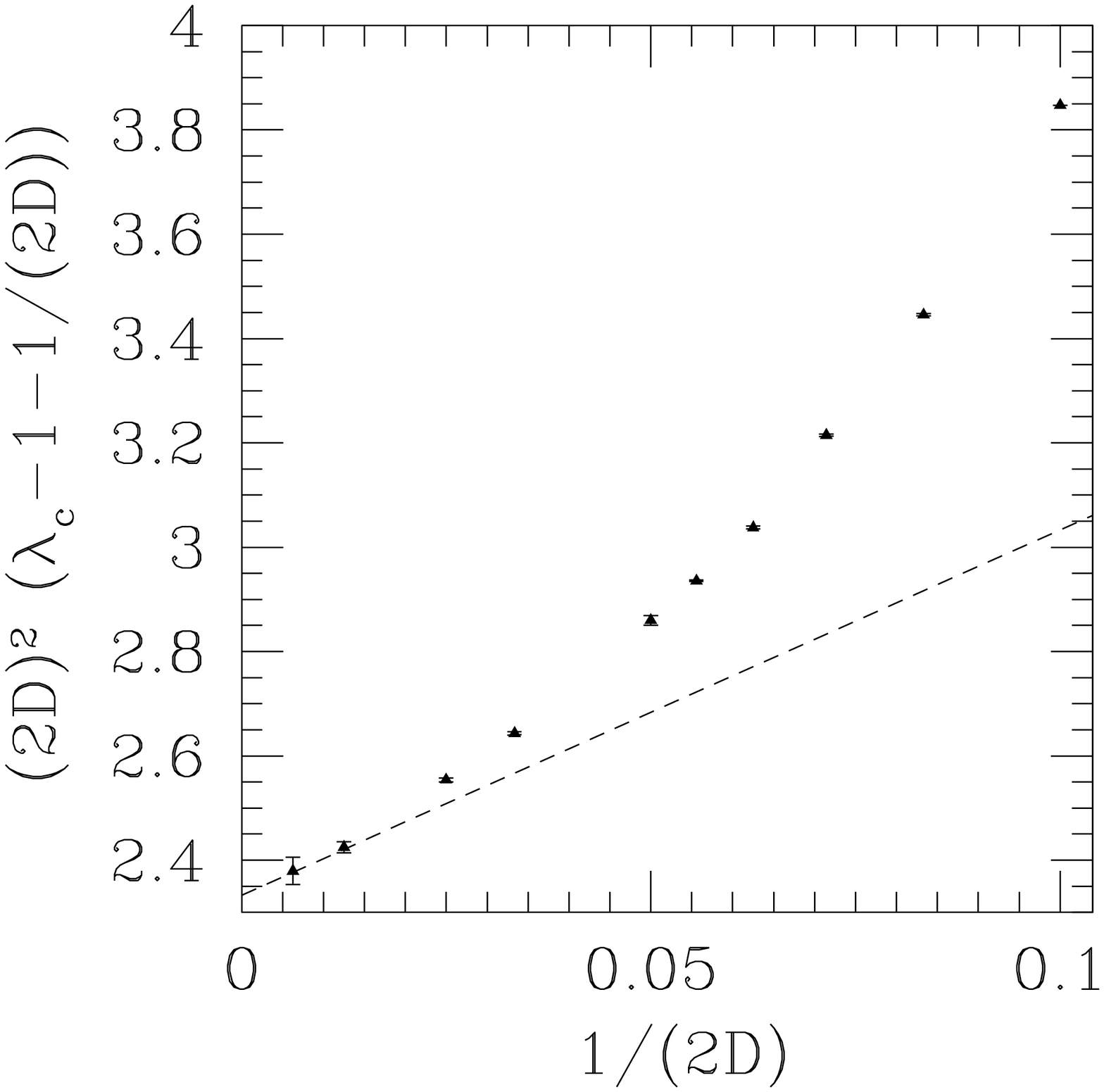,width=7cm}

\caption{Numerical data for the second-order correction in $1/z$ to
$\lambdac(z)$ for the Cayley tree (left panel) and the hypercubic lattice
(right panel). The critical value $\lambdac$ is obtained from numerical
experiments on the conserved contact process (CPP), see main text.  Data
are plotted as $z^2 \,(\lambdac(z)-1-1/z)$ versus $1/z$. The intercept $a$
with the vertical axis ($z \to\infty$) compare very well with the
theoretical predictions $a=4/3$ --- \eqn (\ref{lambdac_arbre_cayley}),
left --- and $7/3$ --- \eqn (\ref{lambdac_hypercubique}), right. Dashed
lines are tentative linear fits (with fixed origin at $1/z=0$) of the 
data, whose slopes should be given by the $O(1/z^3)$ expansion. The 
estimate of $\lambda$ for each $z$ was obtained through an extrapolation 
of data for finite numbers of particles $N$ (up to 4000) to infinite $N$. 
For each value of $N$, we ran 10000 simulations and estimated error bars 
from the statistical fluctuations. The upper and lower values of 
$\lambdac(z;N=\infty)$ were then obtained through a linear fit of 
$\lambdac(z;1/N)$, and plotted at the extremities of the error bars.}
\label{graphes_lambdac}
\end{figure}
\end{center}

Our values of $\rho^*$ are in good agreement with numerical simulations
carried out
on hypercubic lattices in dimensions up to $D=10$ for several values of
$\lambda$ ($=1.5, 2, 3$). Simulating large size lattices in higher 
dimensions would require prohibitive memory space. 
Instead, we have performed simulations of the conserved contact
process (CCP)~\cite{proccont-nconstant-idee}. This is a canonical
counterpart of CP where the number of occupied sites is
kept constant, but $\lambda$ fluctuates. The stationary properties of
both processes are, in the limit of infinite lattice size,
equivalent~\cite{proccont-nconstant-preuve-hilhorst-vanwijland,
proccont-nconstant-preuve-deoliveira}. In particular, simulating CCP with
a (not too large) number $N$ of particles on an almost infinite lattice
(in our simulation, of size $L=2^{32}$) amounts to simulate
CP with a vanishing density. The average value of $\lambda$ in CCP then
coincides with the critical $\lambdac$ in CP. We have been able to simulate
CCP on hypercubic lattices with up to $N=4000$ sites and in dimensions up 
to $D=80$, or up to $z=81$ on the Cayley tree \footnote{$z$ was restricted to 
smaller values in the latter case because simulations on the 
Cayley tree require a slightly more complex computer program to manage the
locations of the particles.}.
Results are displayed in Fig.~\ref{graphes_lambdac}, 
and are in very good agreement with our theoretical predictions for
$\lambdac$~(\ref{lambdac_hypercubique},\ref{lambdac_arbre_cayley})
and previous simulations 
\cite{proccont-nconstant-simulations-sabag-deoliveira}.

\subsubsection{Corrections to the distribution of particle densities}

The analytical resolution of the EM when $\psi (t)\ne 0$ is difficult.
We have restricted ourselves to a first order in $1/z$ expansion around 
the infinite lifetime solutions for which we have an analytical expression 
in the mean-field case. It is convenient to parametrize the real time $t$ 
in terms of $y_0$ introduced in \eqn (\ref{t_de_y_champ_moyen}): $y_0 =
\exp(\psi_0(t))$ where $\psi_0(t)$ is the mean-field expression for 
$\psi(t)$. In the thermodynamic limit $N \to \infty$ with $E \to 0$ 
(infinite lifetime), the $t \to +\infty$ limit translates into the $y_0 
\to \lambda$ limit.

In this setting, we may write $\phi(y_0) = \phi_0(y_0) + \phi_1(y_0)/z +
O(1/z^2)$ and similarly $\psi(y_0) = \psi_0(y_0) + \psi_1(y_0)/z + O(1/z^2)$,
where $\phi_0$ and $\psi_0$ were computed previously for the mean-field
case. From the EM, we derive two coupled first-order linear
differential equations for $\phi_1$ and $\psi_1$:
\begin{equation}
(y_0-1) (\lambda-y_0) \frac{\dd}{\dd y_0}
\left( \begin{array}{c} \phi_1(y_0) \\ \psi_1(y_0) \\ \end{array} \right)
= \frac{1}{y_0}
\left( \begin{array}{cc}
 - (y_0-1)^2 - \lambda + 1  &  \frac{-y_0}{\lambda} (\lambda - y_0) 
 (y_0+1) \\
 2 \lambda (y_0-1)  &  (y_0-1)^2 + \lambda -1
\end{array} \right)
\left( \begin{array}{c} \phi_1(y_0) \\ \psi_1(y_0) \\ \end{array} \right)
+ B(y_0)
\end{equation}
with
\begin{equation}
B(y_0) := \left( \begin{array}{c}
 \lambda^{-2} [-y_0 I_1 + (\lambda - y_0) I_2 + 2 \lambda I_1 I_2 ] \\
 -\lambda^{-1} [2 (y_0-1) I_1 + (\frac{\lambda}{y_0} - 2 (\lambda+1-y_0) )
  I_2 + \frac{2}{y_0} I_1 I_2 ] \end{array} \right)
\end{equation}
and
\begin{eqnarray}
I_1 & := & \left( \frac{\lambda-y_0}{y_0-1} \right)^{\alpha+2}
 \int_1^{y_0} \dd y_1 \frac{\lambda+1-y_1}{y_1-1}
   \left( \frac{y_1-1}{\lambda-y_1} \right)^{\alpha+2} \\
I_2 & := & \left( \frac{y_0-1}{\lambda-y_0} \right)^{\alpha+2}
 \int_{y_0}^\lambda \dd y_1 \frac{y_1}{\lambda-y_1}
   \left( \frac{\lambda-y_1}{y_1-1} \right)^{\alpha+2}
\end{eqnarray}
where $\alpha$ is defined by $\lambda =: 1 + 2/\alpha$. The solutions of
these equations diverge in $y_0=1$ for all initial conditions on
$\phi_1,\psi_1$ in $y_0=1$ except for $\phi_1(1)=-1/(2\lambda^2)$,
$\psi_1(1)=0$ which precisely amounts to setting $\psi(y_0=1)=0$ and
$\phi(y_0=1)=\rho^*$ up to order $1/z^2$. We choose these initial
conditions in the following.

The resolution of the equations for $\phi_1$ and $\psi_1$ can be done in
part analytically and in part numerically. First, we treat the
neighborhood of $y_0$ to characterize exactly the (nondivergent)  
singularity in this point. We have calculated the solutions up to the
order $(y_0-1)^3 \; \ln|y_0-1|$ included, which is sufficient to obtain
the values of $\phi_1 (1+\epsilon), \psi_1 (1+\epsilon)$ slightly off
(below or above) the singularity (with $\epsilon =\pm 0.001$ to $0.003$)
with a good numerical accuracy.  Such an expansion amounts to a short-time
expansion if one starts at $t=0$ from the typical state where the density
of full sites is $\rho^*$ and $y=1$. Then, we start to solve the
differential equations from the value $y_0=1+\epsilon$ using a
Runge-Kutta-Fehlberg procedure. Note that the coefficients of the EM for
$\phi_1$ and $\psi_1$ are rather simple, but the nonvanishing second
member involves, for generic values of $\lambda$, hypergeometric
functions. To simplify the numerical resolution and to make it more
precise, we restricted ourselves to the case $\lambda=1+2/\alpha$ with
$\alpha$ a positive integer, where these hypergeometric functions reduce
to polynomials and logarithms.

This perturbative resolution yields a parametric representation of the
large deviation distribution, $\pi ^*(\rho)= \picm^*(\rho) +
\pi_1^*(\rho)/z + O(1/z^2)$ with parameter $y_0$. More precisely, we have
\begin{eqnarray}
\rho(y_0) &:=& \phi(y_0) = \phi_0(y_0) + \frac{\phi_1(y_0)}z + O(1/z^2)
\quad , \nonumber \\
\pi(y_0) &=& \rho \ln y_0  - S_0(y_0) - \frac{\lambda^2}z \; S_1(y_0) + 
O(1/z^2)\quad
\end{eqnarray} 
where $S_0(y_0) = \ln y_0  - (y_0-1)/\lambda$ and
\begin{equation}
S_1(y_0) = -\frac{1}{\lambda^4} \int_1^{y_0} \dd y_1
 \frac{y_1}{\lambda-y_1}
 \left( \frac{\lambda-y_1}{y_1-1} \right)^{\alpha+2}
 \int_1^{y_1} \dd y_2 \frac{\lambda+1-y_2}{y_2-1}
  \left( \frac{y_2-1}{\lambda-y_2} \right)^{\alpha+2} .
\end{equation}
The resulting curves for $\pi ^*(\rho)$ are presented in 
Fig.~\ref{courbes_pi_de_rho_predites}. Apart from $\pi ^*(\rho^*)=0$ given 
by (\ref{rhomax_hypercubique}, \ref{rhomax_arbre_cayley}) and reached for 
$y_0=1$, another point of interest may be located analytically, namely 
$\pi ^*(\rho=0)$, reached for $y_0=\lambda$. This is related to the 
lifetime of the metastable state, $t_{vac} \sim \exp(-N \pi ^*(\rho=0))$. 
Some values are listed in Table~\ref{table_pi_de_0} for integer $\alpha$.

\begin{table}
\vspace{-.5cm}
\begin{center}
\begin{tabular}{c|c|c}
\hline
$\lambda$ & $\pi(\rho=0)$ & Decimal approximation \\
\hline
3   & $-\ln 3+2/3 + (-53/18 + \pi^2/3)/z$ & $-0.432+0.345/z$ \\
2   & $-\ln 2+1/2 + (-107/36 + \pi^2/3)/z$ & $-0.193+0.318/z$ \\
5/3 & $-\ln (5/3)+2/5 + (-5413/1800 + \pi^2/3)/z$ & $-0.111+0.283/z$ \\
3/2 & $-\ln (3/2)+1/3 + (-1823/600+\pi^2/3)/z$ & $-0.072+0.252/z$ \\
7/5 & $-\ln(7/5)+2/7 + (-270281/88200+\pi^2/3)/z$ & $-0.051+0.225/z$ \\
\hline 
\end{tabular}
\caption{Some analytical values for the $1/z$ expansion of $\pi(\rho=0)$, 
the logarithm of the probability of reaching a configuration with 
vanishing density in the metastable state of the CP on a regular graph of 
coordination number $z$. Analytical calculation can be performed  for 
values of $\lambda=1+2/\alpha$ with $\alpha$ a positive integer only, but 
$\pi (\rho=0)$ can be numerically estimated to order $1/z$ for other 
values of $\lambda$.}
\label{table_pi_de_0}
\end{center}
\end{table}

\begin{center}
\begin{figure}
\epsfig{file=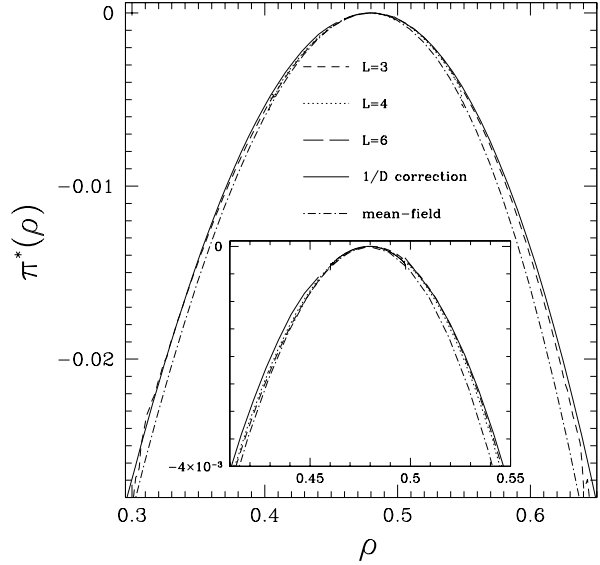,width=8cm}
\caption{Comparison, for $D=6$ and $\lambda=2$,
between the $1/D$ predictions for the large deviation
function $\pi^*(\rho)$ and 
numerical results for CP on hypercubic lattices with periodic boundary 
conditions of linear sizes $L=3$, 4 and 6. All curves were 
horizontally and vertically translated so that their maxima have the same
coordinates $\rho^*,\pi^*=0$. Due to the relatively high dimensionality of 
these systems, the range of values of $\rho$ explored during the 
simulations is very concentrated around $\rho^*$ unless $L$ is small. 
Note that the curves for different lattice sizes, once 
translated, seem to depend only weakly on $L$.
Inset: blow-up of the top region; data compare well with the $1/D$ 
theoretical result (continuous line), and are clearly distinct from 
the mean-field result (dashed-dotted line).}
\label{courbes_pi_de_rho_comparaison}
\end{figure}
\end{center}

To check the accuracy of our perturbative expansion for $\pi^*$, we have
performed simulations of CP on 6-dimensional hypercubic lattices with
periodic boundary conditions. Sizes $N=L^6$ with $L$ ranging from $3$ to
$6$ are large enough that the system gets trapped for a time greater than
the simulation run into the metastable state. We simulated the system
starting from $\rho \ll \rho^*$ or $\rho \gg \rho^*$ until it reached
equilibrium. The equilibration times (expressed as the number of 
elementary steps) was found to correspond to continuous
times~\cite{dickman-reweighting} $t_{eq}$ ranging from 25 to 60.  Then we
run again the simulation for times $t=M\, t_{eq}$ with very large values
of $M$ going from $M=2.5\times 10^5$ for $L=6$ to $M=5\times 10^7$ for
$L=3$, and recorded the histogram of $\rho$ over this time interval. This
gives a very good approximation of the quasistationary distribution
$\pi^*$ since the system was already equilibrated. Numerical results for
$\pi^*(\rho)$ are presented in Fig.~\ref{courbes_pi_de_rho_comparaison}.
The maximum of $\pi^*(\rho)$ vanishes in the thermodynamic limit only, and
the value $\rho^*$ at which this maximum is reached is also subject to
finite-size corrections. To make the comparison easier, we have vertically
and horizontally shifted the experimental curves so that they all reach
zero at the same value of $\rho$. Once this translation was done, the
numerical and theoretical curves are in very good agreement, and are
easily distinguishable from the mean-field curve
(Fig.~\ref{courbes_pi_de_rho_comparaison}). This shows that the curvature
(unaffected by the translations) of $\pi^*(\rho)$ and, hence, the
amplitude of the fluctuations in the metastable state and its lifetime, is
correctly predicted by our $1/z$ calculation.

\section{Conclusion}

In this paper, we have calculated the out-of-equilibrium
distribution of densities of particles in the Contact Process on large 
regular lattices with degree $z$ using
a quantum field-theoretic formulation for the evolution operator.
The calculation is based on a perturbative calculation of the
effective potential for the density $\phi(t)$ and an instantonic
field $\psi (t)$ as function of time $t$. Interestingly, this instantonic
field naturally emerges from the parametrization of the quantum
hard bosonic (or spins-$1/2$) states needed to represent the sets 
of occupation numbers. 

The finite connectivity corrections to the mean-field case
($z=\infty$) lead to the appearance of memory kernels compensating the
loss of information due to the tracking with time $t$
of global fields $\phi(t),\psi(t)$ only. Though calculations may
rapidly become involved, we have been able to show the very good
agreement of the predictions they give with numerical experiments
on the average density of particles and, more generally, the
whole distribution of densities in the out-of-equilibrium metastable
state of CP.

While the use of a quantum formalism was known to be very efficient to
access universal quantities e.g. the decay exponent of the density with
time at criticality, the present study shows that nonuniversal quantities
can be calculated too. We hope that this approach will turn out to be
useful to the analysis of the numerous far--from--equilibrium systems
encountered in physics or related fields. From this point of view, a
remote but promising field of applications could be the analysis of
algorithms in computer science where out-of-equilibrium dynamics over
(discrete) variables abound, and metastability phenomena are present
\cite{walksat-semerjian-monasson, walksat-barthel-hartmann-weigt}.
Hopefully our approach will also permit to complete the average-case
analysis of backtracking algorithms initiated 
in~\cite{dpll-cocco-monasson-prl, coloriage-ein-dor-monasson} through the
systematic control of the non-Markovian effects ignored so far
\footnote{Interestingly, the instantonic field $\psi$ was implicitly
present in the annealed calculation of~\cite{dpll-cocco-monasson-prl, 
coloriage-ein-dor-monasson}, see~\cite{confjapon-backtrack-monasson} for a 
discussion of this point.}.

\begin{acknowledgments} 
We thank A. Georges and M. Sellito for useful discussions, and F. van
Wijland for useful comments about the manuscript. The present work was
partly supported by the French Ministry of Research through the ACI Jeunes
Chercheurs ``Algorithmes d'optimisation et systèmes désordonnés
quantiques''. C.D. acknowledges the hospitality of the Laboratoire de
Physique Théorique in Strasbourg where part of this work was realized. We
thank the ESF SPHINX network for supporting the present work.
\end{acknowledgments}

\appendix

\section{Emergence of memory kernel with hidden degrees of freedom}

Consider two variables $x(t),y(t)$ obeying the Markovian evolution equations
\begin{eqnarray}
\ddt{x}(t) &=& - x(t) + \beta \; y(t) \quad ,\label{trivun} \\
\ddt{y}(t) &=& - \alpha \, y(t) + x(t) \quad ,\label{trideux}
\end{eqnarray}
with initial conditions $x(0)=1$, $y(0)=0$.
This system can be easily solved to give $x$ and $y$ as functions of time
$t$. Assume instead we want to write an evolution equation for
$x$ only. Solving \eqn (\ref{trideux}), and plugging the resulting
$y(t)$ in \eqn (\ref{trivun}), we obtain
\begin{equation}
\ddt{x}(t) = - x(t) + \beta \; \int_{0}^t \ddt'\; e^{-\alpha (t-t')}\;
 x(t') \quad .
\end{equation}
As a result of the existence of a hidden degree of freedom $y$, the 
effective equation on $x$ is not Markovian when $\beta \ne 0$, and includes
a memory kernel whose time constant is that of the $y$ variable.

\bibliography{proccont}

\end{document}